\documentclass[twocolumn]{aastex631}

\usepackage{savesym}
\savesymbol{tablenum}
\usepackage{amsmath,siunitx}
\restoresymbol{SIX}{tablenum}

\usepackage{mathtools}
\usepackage{latexsym}
\usepackage{amssymb}
\usepackage{epsf}
\usepackage{multirow}

\usepackage{color}
\definecolor{lightred}     {rgb}{1.00, 0.35, 0.05}
\definecolor{darkred}      {rgb}{0.70, 0.05, 0.05}
\definecolor{lightblue}    {rgb}{0.05, 0.35, 1.00}
\definecolor{blue}         {rgb}{0.03, 0.25, 0.85}
\definecolor{darkblue}     {rgb}{0.05, 0.05, 0.70}
\definecolor{green}        {rgb}{0.05, 0.65, 0.15}
\definecolor{darkgreen}    {rgb}{0.10, 0.55, 0.15}

\def \kms{km$\,\rm s^{-1}$~}

\def \qb{$\rm Q_b$}

\shorttitle{Non-circular motions of barred galaxies}
\shortauthors{Kim et al.}
\begin{document}

\title{Do strong bars exhibit strong non-circular motions?}

\author[0000-0002-5857-5136]{Taehyun Kim}
%\affiliation{Department of Astronomy and Atmospheric Sciences, Kyungpook National University, Daegu 41566, Republic of Korea}
\affiliation{Department of Astronomy and Atmospheric Sciences, Kyungpook National University, Daegu 41566, Republic of Korea, tkim.astro@gmail.com,  mgp@knu.ac.kr}

\author[0000-0003-1775-2367]{Dimitri A. Gadotti}
\affiliation{Centre for Extragalactic Astronomy, Department of Physics, Durham University, South Road, Durham DH1 3LE, UK}

\author[0000-0003-2779-6793]{Yun Hee Lee}
%\affiliation{Department of Astronomy and Atmospheric Sciences, Kyungpook National University, Daegu 41566, Republic of Korea}
\affiliation{Department of Astronomy and Atmospheric Sciences, Kyungpook National University, Daegu 41566, Republic of Korea, tkim.astro@gmail.com,  mgp@knu.ac.kr}

\author[0000-0003-1045-0702]{Carlos L{\'o}pez-Cob{\'a}}
\affiliation{Institute of Astronomy and Astrophysics, Academia Sinica, No. 1, Section 4, Roosevelt Road, Taipei 10617, Taiwan}

\author[0000-0003-4625-229X]{Woong-Tae Kim}
\affiliation{Department of Physics \& Astronomy, Seoul National University, Seoul 08826, Republic of Korea}
\affiliation{SNU Astronomy Research Center, Seoul National University, Seoul 08826, Republic of Korea}

\author[0000-0002-3560-0781]{Minjin Kim}
\affiliation{Department of Astronomy and Atmospheric Sciences, Kyungpook National University, Daegu 41566, Republic of Korea}

\author[0000-0003-1544-8556]{Myeong-gu Park}
%\affiliation{Department of Astronomy and Atmospheric Sciences, Kyungpook National University, Daegu 41566, Republic of Korea}
\affiliation{Department of Astronomy and Atmospheric Sciences, Kyungpook National University, Daegu 41566, Republic of Korea, tkim.astro@gmail.com,  mgp@knu.ac.kr}

\begin{abstract}
Galactic bars induce characteristic motions deviating from pure circular rotation, known as non-circular motions. As bars are non-axisymmetric structures, stronger bars are expected to show stronger non-circular motions. However, this has not yet been confirmed by observations.
We use a bisymmetric model to account for the stellar kinematics of 14 barred galaxies obtained with the Multi-Unit Spectroscopic Explorer (MUSE) and characterize the degree of bar-driven non-circular motions. For the first time, we find tight relations between the bar strength (bar ellipticity and torque parameter) and the degree of stellar non-circular motions. We also find that bar strength is strongly associated with the stellar radial velocity driven by bars.
Our results imply that stronger bars exhibit stronger non-circular motions.
Non-circular motions beyond the bar are found to be weak, comprising less than 10\% of the strength of the circular motions.
We find that galaxies with a boxy/peanut (B/P) bulge exhibit a higher degree of non-circular motions and higher stellar radial velocity compared to galaxies without a B/P bulge, by $30\sim50\%$.
However, this effect could be attributed to the presence of strong bars in galaxies with a B/P feature in our sample, which would naturally result in higher radial motions, rather than to B/P bulges themselves inducing stronger radial motions.
More observational studies, utilizing both stellar and gaseous kinematics on statistically complete samples, along with numerical studies, are necessary to draw a comprehensive view of the impact that B/P bulges have on bar-driven non-circular motions.
%(the maximum ratio of the amplitude of non-circular motions over circular motions).
\end{abstract}



\section{Introduction}
\label{sec:intro}

% TO DO!
% 최근 연구에 따르면 disk 은하의 비율이 매우 크다 {kartaltepe_23} have shown that 60% of galaxies in the redshift range z=3–6 are disk systems, with about 30% being disky in the interval z = 6–9.)
% As a direct evidence, bar-driven gas inflow rate has been measured for only a few galaxies (\citealt{regan_97,hatchfield_21,sormani_23}).

% ADD Gadotti, Perez, Neuman, , sanchez-blazquez, querejeta, mendez-abrue, adriana caceras, Emsellem, Bland-horthawn, Debattista, etcs

% 막대는 모은하의 현재 모은하의 특성을 결정짖는 중요한 요소 및 앞으로의 진화를 이끄는 주요한 엔진.
Stellar bars are common and important structures that impact the current mass distribution and evolution of their host galaxies. The majority of nearby disk galaxies host large-scale stellar bars. (e.g., \citealt{devaucouleurs_91, eskridge_00, knapen_00, menendez_delmestre_07, mendez-abreu_10, nair_10a, masters_11, buta_15}). 
Although the fraction of barred galaxies declines with increasing redshift (e.g., \citealt{abraham_99, sheth_08, cameron_10, melvin_14}), recent studies have found barred galaxies at z$\sim$3 (e.g., \citealt{guo_23, costantin_23, leconte_24, Amvrosiadis_24xx}), and find some evidence of bars even at z$>$4 (e.g., \citealt{hodge_19, smail_23, tsukui_24}). This indicates that bars have influenced their host galaxies over a significant period of their lifetimes.

Bars are non-axisymmetric structures that drive radial angular momentum transfer, resulting in gas inflow toward the center and the accumulation of gas in the central regions of galaxies (e.g., \citealt{roberts_79, athanassoula_92a, sellwood_93, sakamoto_99, sheth_00, sheth_02, jogee_05, kim_w_12a, sormani_15b}).
The gas accumulation triggers star formation and facilitates the formation of various stellar structures, including nuclear disks, nuclear rings, inner bars, and nuclear spiral arms. Additionally, this gas accumulation may potentially fuel AGN activity (e.g., \citealt{knapen_95, knapen_02, ho_97, maciejewski_04b, sheth_05, comeron_10, vandeven_10, coelho_11, ellison_11, delorenzo-caceres_12, gadotti_15, gadotti_20, ridley_17, kim_17, neumann_20, desafreitas_23a, desafreitas_23b, kolcu_23, zee_23, garland_24}).
Barred galaxies experience the vertical buckling instability, which leads to vertical growth within the inner part of the bar and the formation of boxy/peanut (B/P) or X-shaped bulges (e.g., \citealt{combes_81, desouza_87, chung_04, bureau_05, debattista_05, laurikainen_05, laurikainen_14, laurikainen_17, martinez_valpuesta_06, mendez-abreu_08, erwin_13, erwin_17, athanassoula_15, laurikainen_05, laurikainen_14, laurikainen_17, fragkoudi_15, fragkoudi_17b, fragkoudi_20, kruk_19}). 
Vertical resonances have also been found to contribute to the formation of B/P features. (e.g., \citealt{combes_81, quillen_02, debattista_06, sellwood_20, li_23}).
Bars push the gas between the corotation radius and the outer Lindblad resonance radius outwards (\citealt{combes_85, combes_08_gas}). In consequence, bars are closely associated with the presence of the outer ring where the outer Lindblad resonance is expected (e.g., \citealt{schwarz_81, buta_96, buta_03, romero_gomez_06, athanassoula_09b, buta_15}). Disk breaks in the radial surface brightness profiles are frequently found at the outer ring in barred galaxies (\citealt{erwin_08, munoz_mateos_13, kim_14}).
%radial surface brightness profiles show disk breaks near the outer ring in barred galaxies (\citealt{erwin_08, munoz_mateos_13, kim_14}).
%many disk truncations are found near the outer ring in barred galaxies (\citealt{erwin_08, munoz_mateos_13, kim_14}).
Thus, bars play a defining role in shaping the complex morphology and the mass distribution of galaxies.

% Impact of bars on gaseous & stellar velocity fields.
Such distinctive mass distribution of barred galaxies naturally leads to show a characteristic dynamics, such as oval distortions appeared as ``S-shaped'' velocity fields in both gaseous and stellar kinematics.
These velocity fields cannot be explained with pure circular rotation but exhibit non-circular, streaming motions of gas (e.g., \citealt{bosma_78, pence_81, kormendy_83, pence_84, weliachew_88, fathi_05,spekkens_07}).

Bars have a significant impact on both the mass distribution and kinematics of stellar disks (e.g., \citealt{ghosh_23b}). 
In the Milky Way, the bar leaves imprints on the stellar density structures, particularly near its major resonances (\citealt{khoperskov_20a}). Bars also impact stellar kinematics, especially in the inner regions (e.g., \citealt{seidel_15b, gadotti_20, walo-martin_22}). Bars induce stellar migration due to the resonance overlap of the bar and spiral structure (\citealt{minchev_10, ghosh_23b}). 
In cosmological zoom-in simulations, bars are found to enhance the radial velocity dispersion (\citealt{Pinna_18}), and drive a low velocity dispersion along the bar in $\sigma_z$ and $\sigma_{\phi}$, thus leading to a significant velocity dispersion differences between the bar major and minor axis (\citealt{walo-martin_22}).

Recently, detailed estimations of radial velocity driven by bars have become feasible with the availability of high-resolution two-dimensional velocity maps (\citealt{holmes_15, salak_19, lopez-coba_22, hogarth_24, digiorgio_zanger_24xx}).
While non-circular motions can also originate from triaxial dark halos, spiral arms, warps, or galaxy lopsidedness (\citealt{bosma_78, begeman_87, hayashi_04, stark_18, oman_19}), bars are the primary driver of non-circular motions in barred galaxies (\citealt{spekkens_07, holmes_15, stark_18, lopez-coba_22}).

% NC motions은 여러 방법으로 측정가능. (Fourier series & bisymmetric, removing Cir-motions)
% Diskfit algorithm :(Spekkens & Sellwood 2007; Sellwood & Sanchez ´ 2010; Sellwood & Spekkens 2015)
% Fourier series
Non-circular motions in barred galaxies have been modeled using either Fourier series (e.g., \citealt{schoenmakers_97, salak_19}) or bisymmetric models (e.g., \citealt{spekkens_07, hallenbeck_14, holmes_15, lopez-coba_22, digiorgio_zanger_24xx}).
Using Fourier analysis on CO(1-0) velocity maps, \citet{salak_19} find that barred galaxies exhibit a higher ratio of non-circular to circular velocities of molecular gas by a factor of 1.5 -- 2, compared to non-barred galaxies within the bar radius.  
%They also find that the ratio of non-circular to circular motions peaks at a radius of $0.3\times \rm R_{bar}$ where $\rm R_{bar}$ is the bar radius.
Using the bisymmetric model on VLT/MUSE datasets, \citet{lopez-coba_22} modeled non-circular motions in the stellar and $\rm H\alpha$ velocity maps of barred galaxies.
They find that the kinematic position angle (PA) of the non-circular motions closely matches the photometric PA of the bar. This implies that the non-circular motions are caused by bars. 
They demonstrate that both the bisymmetric model and Fourier series analysis effectively explain the oval distortions observed in the velocity maps of barred galaxies.
Additionally, they show that when galaxies are gas-rich and observed with enough signal-to-noise ratio (S/N) in H$\alpha$, non-circular motions are also observed in H$\alpha$ velocity maps, but with larger amplitudes compared to the ones from the stellar velocity maps (See also \citealt{digiorgio_zanger_24xx}).
Using a simplified bisymmetric model on ALMA CO(1$-$0) velocity maps, \citet{hogarth_24} find that barred galaxies are more likely to exhibit large-scale radial gas motions compared to their unbarred counterparts.

% From circular motion modeling
Instead of modeling complex non-circular motions directly, \citet{erroz-ferrer_15} model the circular motions of galaxies from the galaxy rotation curve and then create the residual velocity map by subtracting the model of circular motions from the galaxy rotation map. Using the cumulative distribution function on the residual velocity map, they estimate the amplitude of non-circular motions and find that non-circular motions are weakly related to the star formation rate of the bar region, such that star forming bars exhibit slightly higher non-circular motions.

% Santi's 15 paper:
% One of the aims of this kinematical study is to analyse the influence of the galaxies’ main structural features on their kinematics. In other words, we want to study the kinematical footprints of the components of the galaxies, traced for instance by the non-circular motions. Deviations from pure circular motion have been widely studied previously in the literature

%(check: whether there's a aligned bar horizontally/perpendicularly)

%Radial gas inflow는 bar strength가 클수록 강할 것으로 예측 
As stated above, the majority of non-circular motions in barred galaxies are bar-driven (e.g., \citealt{bosma_78, pence_88, spekkens_07, stark_18, lopez-coba_22}). 
Then what properties of the bar control these bar-driven non-circular motions? Bar-driven gas inflow mainly occurs along the bar dust lane (e.g., \citealt{athanassoula_92b, englmaier_97, kim_w_12a, emsellem_15, sormani_15b, sormani_23, seo_19}).
In the numerical simulations of \citet{athanassoula_92b}, substantial gas inflow is found when there are strong shocks at the bar dust lane.
One piece of direct evidence of shock at the dust lane is a velocity jump across the dust lane (\citealt{athanassoula_92b}). Observational studies find velocity jumps of 100 -- 200 \kms at the dust lane in various barred galaxies (\citealt{pence_84, reynaud_98, laine_99, mundell_99, zurita_04, zanmar_sanchez_08, feng_22, kim_24}). 
%위와 붙여서
\citet{athanassoula_92b} find that the velocity jump increases as a function of the bar axis ratio and the quadrupole moment of the bar (see their figure 12). 
This implies that strong bars tend to have strong shocks at the bar dust lane, thereby driving more gas inflow. 
\citet{regan_04} also find that strong bars that form nuclear rings induce significant gas inflow, whereas weak bars have almost no effect on the radial gas distribution.
Therefore, bar-driven non-circular motions are expected to be closely related to bar strength.
The main goal of this paper is to examine whether stronger bars show a higher degree of non-circular motions using observational data. 

Another property of a bar that affects gas inflow is the presence of a B/P bulge. \citet{fragkoudi_16} find that if a galaxy has a strong B/P bulge, the gas inflow decreases by a factor of 10 compared to barred galaxies without a B/P bulge. We also aim to investigate the impact of hosting a B/P bulge on non-circular motions.

The paper is organized as follows.
We describe the data used and data analysis in Section 2. 
In Section 3, we investigate the relation between the degree of non-circular motions and the bar strength. 
Section 4 explores the impact of having B/P bulges on the degree of non-circular motions.
In Section 5, we discuss the limitations of adopting the bisymmetric model to describe non-circular motions.
Finally, we summarize and conclude in Section 6.

\section{Data and data analysis} \label{sec:data}

\begin{figure*}[th!]
\includegraphics[width=\textwidth]{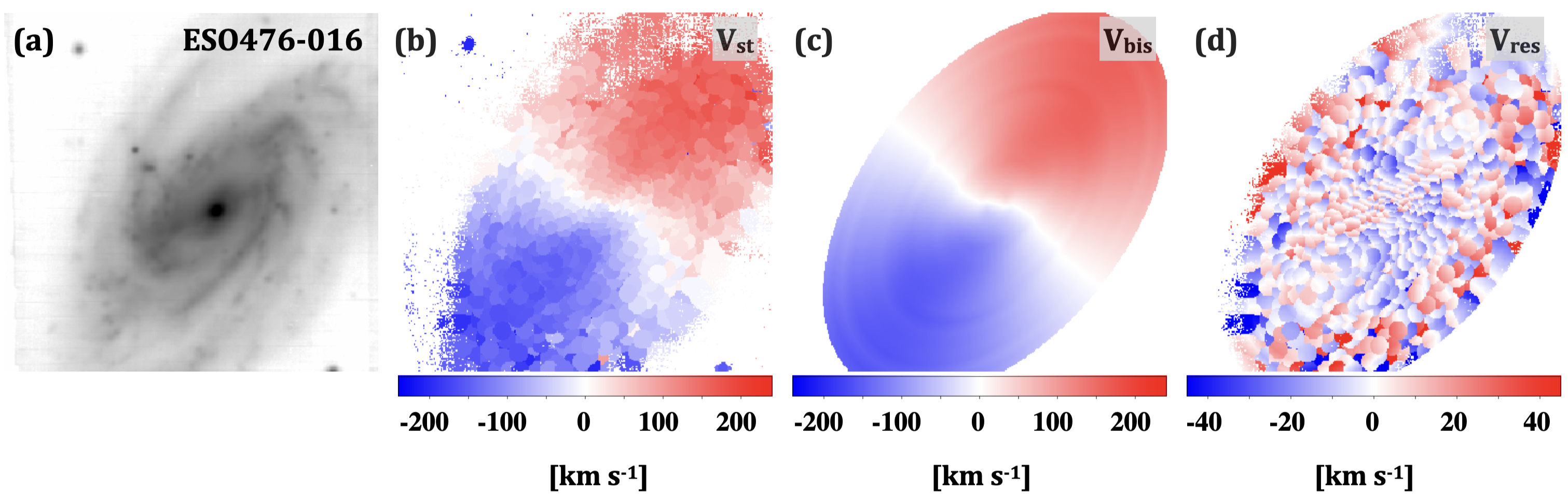}
\caption{ESO476-016, (a): Optical r-band image constructed from the MUSE datacube, (b): stellar velocity map ($V_{\rm st}$), (c): bisymmetric model ($V_{\rm bis}$), and (d): residual velocity ($V_{\rm res}$ = $V_{\rm st} -V_{\rm bis}$). (b),(c) and (d) are taken from \citet{lopez-coba_22} and reconstructed.
\label{fig:bimod_set4}
}
\end{figure*}

\subsection{Non-circular motions}
We utilize data from \citet{lopez-coba_22}, which selected barred galaxies from the AMUSING++ survey (\citealt{galbany_16, lopez-coba_20}). 
Barred galaxies were selected based on the following criteria: (i) the bar is clearly visible in gri images, (ii) the galaxy is non-interacting, (iii) observations were made under optimal atmospheric conditions, (iv) the bar length is well resolved in the stellar velocity map, and (v) the bar is fully covered within the MUSE field of view.
The AMUSING++ sample continues to grow, and upon re-evaluating the sample, we find that IC 2151 meets the selection criteria outlined by \citet{lopez-coba_22}, thus we add IC2151. Our sample consist of 14 galaxies and the basic information is presented in Table~\ref{tab:sample}.
The galaxies are observed with the Multi-Unit Spectroscopic Explorer (MUSE, \citealt{bacon_10}), which is installed at the Very Large Telescope (VLT). 
The MUSE integral field unit dataset contains $\sim$90,000 (300$\times$300) spaxels per observation. Each spaxel covers 4800 \r{A} to 9300 \r{A} with a resolving power of 1770 at 4800 \r{A} and 3590 at 9300 \r{A} and with spatial sampling of 0.2" per spaxel. 
The field of view for each object is 1'$\times$1'.

%\textcolor{blue}{Details on obtaining stellar kinematics are described in \citet{lopez-coba_20}. In summary, the sample data were reduced using \texttt{Pipe3D} (\citealt{sanchez_16}). Stellar population libraries are taken from the MIUSCAT templates (\citealt{vazdekis_12}), and the stellar velocity is derived under the assumption that all stellar populations move at the same velocity, with a velocity dispersion following a Gaussian function (\citealt{sanchez_16}).}

Details on obtaining stellar kinematics are described in \citet{lopez-coba_20}. In brief, the data were analyzed using the \texttt{Pipe3D} pipeline (\citealt{sanchez_16}).
\texttt{Pipe3D} performs a decomposition of the observed spectra into multiple single stellar populations (SSP), with different ages and metallicities. The MIUSCAT templates (\citealt{vazdekis_12}) are adopted. In order to increase the S/N of the stellar continuum, \citet{lopez-coba_22} performed a segmentation procedure of the spectra around the MUSE V-band. This procedure coadds spectra until reaching an approximately constant S/N ($\sim$50), at the cost of degrading the original spatial resolution of the data. Then the stellar velocity is derived in these coadded spectra under the assumption that the stellar populations move at the same velocity and velocity dispersion. When obtaining stellar kinematics, emission lines are masked.
We refer to \citet{lopez-coba_20} and \citet{sanchez_16} for a thorough description of this analysis.
In addition to the samples from \citet{lopez-coba_22}, we include IC 2151 which is taken from the AMUSING++(\citealt{lopez-coba_20}), and analysed using the same method as the other sample galaxies. 
%IC 0004 is excluded in our sample as we obtain large errors in Vrad & Vtan in the following analysis.

%% Maybe later add PA of the disk, bar, Qb, emax. 
\begin{deluxetable*}{lccccccclcc}
\tablecaption{Basic properties of sample galaxies. \label{tab:sample}}
\tabletypesize{\scriptsize}
\tablewidth{0pt}
\tablehead{ \colhead{Galaxy} & \colhead{Log($M_{*}$)} & \colhead{ PA$_{\rm Bar}$} & \colhead{ PA$_{\rm Disk}$} & \colhead{emax} &  \colhead{\qb} & \colhead{$A_2$} & \colhead{$R_{\rm bar}$} &\colhead{B/P} & \colhead{$R_{\rm BP}$} & \colhead{$R_{\rm BP_{dp}}$}
\\
 \colhead{} & \colhead{[$M_{\odot}$]} & \colhead{ [deg]} & \colhead{[deg]} & \colhead{}& \colhead{}& \colhead{[arcsec]}& \colhead{[arcsec]} &  \colhead{} & \colhead{[arcsec]} & \colhead{[arcsec]} }
\decimalcolnumbers
\startdata
        ESO 018 - G018  & 10.5 &  53  & 281 & 0.497 & 0.158 & 0.18  & 5.24   & --          & -- & -- \\ 
        ESO 325 - G043  & 11.1 &  28  & 109 & 0.454 & 0.196 & 0.233 & 6.55   & --          & -- & -- \\ 
        ESO 476 - G016  & 10.8 &  114 & 321 & 0.617 & 0.333 & 0.168 & 15.98  &  B/P     & 4.17  & 4.87 \\ 
        IC 2151$^{(a)}$   & 10.7 &  76  & 100 & 0.696 & 0.576  &  0.290  & 7.074  & -- $^{(b)}$ & -- & --  \\
        IC 2160         & 10.6 &  144 & 105 & 0.620 & 0.512  &  0.311  & 19.91  & B/P & 4.60 & 5.54\\ 
        NGC 0289        & 10.5 &  120 & 128 & 0.547 & 0.226  &  0.161  & 20.96  & B/P   & 7.69  & 7.77\\ 
        NGC 0692        & 11.2 &  117 & 259 & 0.450 & 0.280  &  0.200  & 11.27  & --     & -- & -- \\ 
        NGC 1591        & 10.4 &   47 &  23 & 0.609 & 0.392  &  0.197  & 10.48  & --     & -- & -- \\ 
        NGC 3464        & 10.6 &  111 & 109 & 0.575 & 0.263  &  0.105  & 22.27  & --     & -- & -- \\ 
        NGC 5339        & 10.4 &   83 & 208 & 0.573 & 0.451  &  0.236  & 25.68  & B/P   & 10.03  & 10.81  \\ 
        NGC 6947        & 10.9 &    4 & 238 & 0.669 & 0.492  &  0.228  & 28.82  & B/P & 7.30   & 9.66 \\ 
        NGC 7780        & 10.5 &  160 & 188 & 0.508 & 0.240  &  0.210  & 18.08  & B/P &  6.47  & 8.79\\ 
        PGC 055442      & 10.9 &   39 & 193 & 0.291 & 0.094  &  0.108  & 5.50   & --      & -- & -- \\ 
        UGC 03634       & 11.0 &    1 & 120 & 0.477 & 0.220  &  0.192  & 12.84  & --   & --  & -- \\ 
\enddata
\tablecomments{(1): Galaxy name, (2): stellar mass , (3): photometric PA of the bar, (4): photometric PA of the disk, (5): maximum ellipticity near the bar end, (6): \qb, (7): $A_2$, (8): bar radius obtained at the maximum ellipticity which measured on deprojected images, (9): type of bulge either disky or B/P bulge, decided in the Sec.~\ref{sec:anal_bp}, (10): the radius of B/P bulges where B4 parameter reach its minimum inside the bar. $R_{\rm BP}$ is obtained on projected images, and (11): the deprojected $R_{\rm BP}$. $R_{\rm BP}$ is deprojected analytically.
(2) -- (4) are from \citet{lopez-coba_22}.
$^{(a)}$: IC 2151 data is taken from the AMUSING++ survey (\citealt{lopez-coba_20}). $^{(b)}$: IC 2151 does not have a prominent bulge.}
\end{deluxetable*}

If we assume that particles follow circular orbits in a galaxy, the simple circular velocity along the line-of-sight direction can be modeled as,

\begin{equation}
    V_{cir} = V_{sys} + V_{t} \sin \textit{i} \; \cos \theta,
\end{equation}
\label{eq:vcir}

where $V_{sys}$ is the systemic velocity of the galaxy, $V_{t}$ is the circular rotation velocity, \textit{i} is the inclination of the galaxy and $\theta$ is the azimuthal angle in the disk plane relative to the projected major axis.

%We made use of the \texttt{XookSuut (XS)} code (\citealt{lopez-coba_21}) to model circular and non-circular motions using velocity maps of galaxies. 
%To model circular and non-circular motions using velocity maps of galaxies, the \texttt{XookSuut (XS)} code is employed. \texttt{XS} is a Python tool which combines \texttt{DiskFit} (\citealt{spekkens_07}) models and Fourier analysis to extract non–circular rotation patterns from velocity fields. We retrieve the bisymmetric model of our sample galaxies from \citet{lopez-coba_22}. The bisymmetric model describes an oval distortion generated by stellar bars on the velocity field and can be equated as follows,

In order to derive non-circular motions of barred galaxies, non-axisymmetric components are added in the model fitting in addition to Eq. 1. The bisymmetric model (\citealt{spekkens_07}) describes an oval distortion generated by stellar bars on the velocity field and can be equated as follows,

\begin{equation}
\begin{split}
V_{bis,model} = V_{sys} + \sin \textit{i} \; (V_t \cos \theta - V_{2,t} \cos 2(\theta_{b}) \cos \theta \\
- V_{2,r}  \sin2(\theta_{b})  \sin \theta ).
\end{split}
\label{eq:bis}
\end{equation}

The position angle of the major axis of the bar is $\phi_{b}$, which is measured from the projected major axis of the disk. The angle relative to the bar axis is then $\theta_{b} = \theta - \phi_{b}$. All the angles, $\theta$, $\theta_{b}$ and $\phi_{bar}$, are measured on the disk plane. 
$V_{2,r}$ represents the radial velocity and $V_{2,t}$ represents the tangential velocity produced by the oval distortions (e.g., bars) and both are function of the galactocentric distance. The schematic geometry of the disk plane can be seen in Figure 1 of \citet{spekkens_07}.

We utilize the results from the bisymmetric model fitting performed by  \citet{lopez-coba_22}. To simultaneously model circular and non-circular motions using galaxy velocity maps, \citet{lopez-coba_22} employed the \texttt{XookSuut (XS)} code (\citealt{lopez_coba_xs24}). XS is a Python-based tool that integrates DiskFit (\citealt{spekkens_07}) with Fourier analysis to extract non-circular rotation patterns from velocity fields. Through the application of XS, $V_{sys}, i, \phi_{b}$, radial profiles of $V_t, V_{2,r}$ and $V_{2,t}$ from Equation \ref{eq:bis} can be obtained, and we retrieve these value from \citet{lopez-coba_22}.

H$\alpha$ kinematics provide essential information about the gaseous kinematics (e.g., \citealt{daigle_06, erroz-ferrer_15}). However, due to the low S/N of H$\alpha$ in some parts of our sample galaxies, the bisymmetric models on H$\alpha$ are only available for less than half of the sample (6 out of 14), and the H$\alpha$ velocity maps are patchy, especially around the bar. Therefore, we focus on stellar kinematics for further analysis.

We present the r-band image, the stellar velocity map, the bisymmetric model, and the velocity residual map of ESO476-016 in Figure~\ref{fig:bimod_set4}, as an example. 
The radial profiles of $V_{2,r}$ (radial velocity, $V_{\rm rad}$), $V_{2,t}$ (tangential velocity, $V_{\rm tan}$) and $V_{t}$ (circular velocity, $V_{\rm cir}$) from the bisymmetric model are plotted in Figure~\ref{fig:radial_mod_strength}. Radial profiles of the bisymmetric models for the other sample galaxies are presented in the Appendix.

To quantify the non-circular motions, we employ the amplitude of the bisymmetric non-circular motions, $A_{bis}$ (\citealt{lopez-coba_22}), using the amplitude of the two parameters, the radial velocity ($V_{2,r}$) and the tangential velocity ($V_{2,t}$) as,

\begin{equation}
    A_{bis} (r) = \sqrt{V_{2,r}^2(r) + V_{2,t}^2(r)}.
\end{equation}

Then, we estimate the fraction of the non-circular motions divided by the circular rotation velocity at each radius by
\begin{equation}
    f_{\rm NC} (r) = \frac{A_{bis}(r)}{V_{t}(r)},
\label{eq:fnc}
\end{equation}
where $V_{t}(r)$ is the circular velocity defined in Equation \ref{eq:vcir}. 
We plot $f_{\rm NC} (r)$ in Figure \ref{fig:radial_mod_strength}(b).

\subsection{Bar strength}
We estimate bar strength using r-band images from the Dark Energy Spectroscopic Instrument (DESI) Legacy Imaging Surveys Data Release 10 (\citealt{dey_19}), which combines data from the Dark Energy Camera Legacy Survey (DECaLS, \citealt{flaugher_15}), the Beijing–Arizona Sky Survey  (BASS, \citealt{zou_17_bass}), and the Mayall z-band Legacy Survey (MzLS, \citealt{dey_16, dey_19}). When DESI images are not available (e.g., UGC03634), we obtain the data from the Panoramic Survey Telescope and Rapid Response System (Pan-STARRS, \citealt{chambers_16, flewelling_20}). 

The pixel scales of the images are 0.262"/pix (DESI) and 0.258"/pix (Pan-STARRS). The median FWHM of the r-band DESI images is $1\farcs2$ -- $1\farcs5$. The median 5$\sigma$ detection limit in the r-band is 23.3 AB mag for DeCALS and 22.9 AB mag for BASS (\citealt{dey_19}).
Bar strengths are estimated on deprojected images. 
%The median FWHM of the of the r-band DESI images is 1.2" for the DECaLS and 1.5" for the BASS.

%% The "ht!" tells LaTeX to put the figure "here" first, at the "top" next
%% and to override the normal way of calculating a float position
\begin{figure}[ht!]
\includegraphics[width=0.5\textwidth]{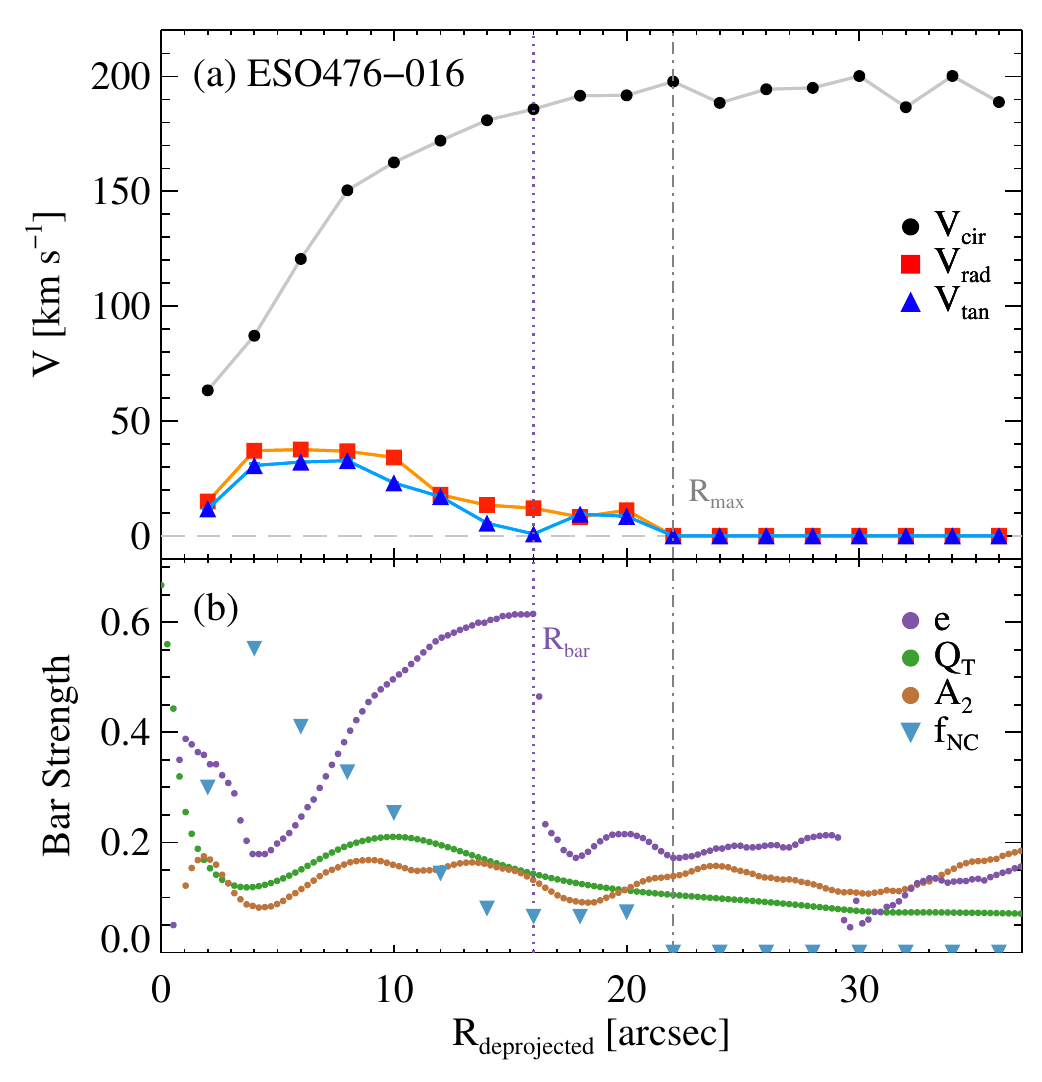}
\includegraphics[width=0.5\textwidth]{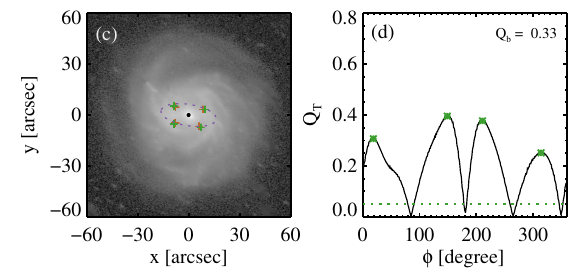}
\caption{(a): The radial profiles of $V_{\rm cir}, V_{\rm rad}$, and $V_{\rm tan}$ for ESO476-016, obtained by fitting the velocity fields with the bisymmetric model (\citealt{spekkens_07}). 
(b): The radial profiles of ellipticity (e), \qb, and $A_2$. 
Vertical dotted line denotes the bar radius ($R_{\rm bar}$) estimated from the maximum of the bar ellipticity, vertical dashed-dotted line stands for $R_{\max}$ where the non-circular motions ($V_{\rm rad}$ and $V_{\rm tan}$) become zero for more than two consecutive points.
(c): Deprojected image of ESO476-016, overlaid with $R_{\rm bar}$ in a dashed ellipse. Green `+' points denote positions where \qb~is estimated, bronze `+' points denote positions where $A_2$ is estimated.
(d): The azimuthal profile of $Q_T$ at the radius where the radial profile of $Q_T$ reach its maximum in (c). The local maxima of the azimuthal profile of $Q_T$ are denoted by green asterisks. The dotted line at $Q_T = $0.05 marks the threshold to count the number of peaks associated with the bar. \qb~is the average of the four local maxima.
\label{fig:radial_mod_strength}
}
\end{figure}

Bar strength is not a uniquely defined quantity (\citealt{athanassoula_13}) and can be measured using various methods, such as ellipticity, the non-axisymmetric torque parameter, and the amplitude of the Fourier components (e.g., \citealt{gadotti_07, gadotti_11, garcia_gomez_17, lee_20}).

% Ellipticity
In barred galaxies, the radial profile of ellipticity increases along the bar radius and typically peaks near the end of the bar, as shown in Figure~\ref{fig:radial_mod_strength}(b). The maximum ellipticity near the bar end, referred to as the bar ellipticity, can serve as an indicator of bar strength. 
While this measure may underestimate the true ellipticity of the bar (e.g., \citealt{gadotti_08}) due to the coexistence of the bar with the disk and bulge (if present), it has been extensively used as a measure of bar strength (e.g., \citealt{martin_95, abraham_99, aguerri_99, marinova_07}) because it is straightforward to obtain and model-independent.

% --- Qb
Originated from numerical simulations of disk galaxies, the bar torque (\qb) is introduced by \citet{combes_81}. It is defined as the maximum value of the ratio of the tangential force to the mean axisymmetric radial force. It has been widely used to measure bar strength in numerous studies (e.g., \citealt{quillen_94, salo_99, buta_01, block_01, laurikainen_02_2mass, diaz-garcia_16a, garcia_gomez_17, lee_20}).

To estimate \qb, we follow the procedures described in \citet{lee_20}. The gravitational potential map ($\Phi$) is constructed from r-band images from DESI or Pan-STARRS, assuming a constant mass-to-light ratio (M/L) throughout the disk \citep{quillen_94}. The three-dimensional mass density is estimated from the two-dimensional mass map by convolving it with the vertical density profile prescribe as follows. The vertical profiles of the disks are assumed to be exponential. To estimate the vertical scale height of disk ($h_z$), we applied the average empirical correlations between the radial scale length, $h_r$, and the vertical scale height, $h_z$, depending on the Hubble type. The average ratio of $h_r/h_z$ is taken from \citet{laurikainen_04b}, and the Hubble type of each galaxy is from Hyperleda\footnote{http://leda.univ-lyon1.fr/}(\citealt{makarov_14}).
The average ratio of $h_r/h_z$ is set to 4 for T $\leq$ 1, 5 for 1 $<$ T $\leq$ 4, 9 for T $>$ 9, following the value from \citet{laurikainen_04b}.
From the gravitational potential map in the polar coordinated, $\Phi(r,\phi)$, we can define the mean radial force ($\langle F_{\rm R}(\rm r) \rangle$), tangential force ($F_{T}$), and the ratio of the two ($Q_{T}$) as,
% 수식 넣기!
\begin{equation}
    \langle F_{\rm R}(\rm r) \rangle = \frac{d \Phi_{0}(r)}{dr},
    \end{equation} 

\begin{equation}
     F_{T}(r,\phi) = \Bigg| \frac{1}{r} \frac{\partial \Phi(r,\phi)}{\partial \phi} \Bigg|,
\end{equation} 

and 
%\begin{equation}
%     Q_{T}(r)   \equiv  \frac{F_{T}^{\rm max}(r)}{ \langle F_{R}(r)\rangle}
%\end{equation} 

\begin{equation}
 Q_{T}(r,\phi) = \frac{F_{T}(r,\phi)}{\langle F_{R}(r) \rangle}
\end{equation} 

where $\Phi_{0}$ is the m$=$0 Fourier component of the gravitational potential (\citealt{combes_81, buta_01}). We plot the radial profile of $Q_{T}$ in Figure~\ref{fig:radial_mod_strength}(b).
We then define the bar strength \qb~as the average of $Q_{T}$ in the polar coordinate system as follows. 

\begin{equation}
 Q_{b} \equiv \frac{1}{p} \sum_{i=1}^p Q_{T,i},
\end{equation} 

where $Q_{T,i}$ is the maximum $Q_{T}(r,\phi)$ at each peak on the azimuthal profile (Figure~\ref{fig:radial_mod_strength}d) and $p$ represents the number of peaks. Normally for barred galaxies, p is four, but it can vary if the bar is not strong enough or due to the presence of strong spiral arms.

% Caveat
%While \qb~is an widely used bar strength indicator, \citet{regan_04} show that potentials with nearly identical values of \qb~can lead to have substantially different orbital morphologies and gas inflow rate. For example, they argue that nearly identical values of \qb, the mass inflow can vary by a factor of $\sim$ 8. Therefore, we also measure another bar strength parameter as the following.
%They claim that \qb is not able to represent the conditions needed for various bar orbit family transitions.
%\qb is degenerate between the bar quadrupole moment and bar axis ratio.

%We estimate the maximum amplitude of the relative m = 2 component (e.g., \citealt{athanassoula_13}). 

% --- A2 
%Bar strength can also be obtained from the Fourier decomposition on two dimensional mass distribution as the sum of the two Fourier components, $a_m(R)$ and $b_m(R)$, which are defined as follows (e.g., \citealt{ohta_90, athanassoula_13, lee_19}).

%\begin{equation}
%a_m(R) = \sum_{i}{m_i} \cos(m\theta_{i}),
%\end{equation}
%\begin{equation}
%b_m(R)= \sum_{i}{m_i} \sin(m\theta_{i}),
%\end{equation}

%where $R$ is the cylindrical radius, $m_i$ is the mass , $\theta_i$ is the azimuthal angle of the $i$-th bin at the radius R, respectively.vThe maximum amplitude of the $m=2$ component ($A_2$), is defined as
%\begin{equation}
%A_2 \equiv \rm{max} \left(\frac{\sqrt{a_2^2+b_2^2}}{a_0}\right).
%\end{equation}.
%We plot profiles of ellipticity, \qb~ and $A_2$ in the lower panel of Fig. \ref{fig:radial_mod_strength}.

Bar strength can also be obtained from Fourier decomposition. The azimuthal profiles of each isophote ($I(r,\theta)$) are defined as follows (e.g., \citealt{ohta_90, aguerri_98, lee_19}, See also \citealt{athanassoula_13}).

%\begin{equation}
%I(r,\theta) = A_0 + \sum_{m=1}^{\infty} [a_m(r) \cos m\theta + b_m(r) \sin m\theta ],
%\end{equation}
\begin{equation}
I(r,\theta) = I_0(r) + \sum_{m=1}^{\infty} [a_m(r) \sin m\theta + b_m(r) \cos m\theta ],
\end{equation}
where

%\begin{equation}
%\begin{aligned}
%a_m(r) = \frac{1}{\pi} \int_0^{2\pi} I(r,\theta) \cos m\theta \, d\theta 
%\\
%b_m(r) = \frac{1}{\pi} \int_0^{2\pi} I(r,\theta) \sin m\theta \, d\theta.
%\end{aligned}
%\end{equation}

\begin{equation}
\begin{aligned}
a_m(r) = \frac{1}{\pi} \int_0^{2\pi} I(r,\theta) \sin m\theta \, d\theta 
\\
b_m(r) = \frac{1}{\pi} \int_0^{2\pi} I(r,\theta) \cos m\theta \, d\theta. 
\end{aligned}
\label{eg:fourier_eq2}
\end{equation}

Note that the cosine and sine terms are switched to follow the \texttt{IRAF} \texttt{ellipse} convention for obtaining the boxiness parameter, which will be explored in the next subsection.
The Fourier amplitude of the $m$-th component is defined as,
\begin{equation}
%A_2 (r) = \frac{I_2(r)}{I_0(r)} = \left(\frac{\sqrt{a_2(r)^2 + b_2(r)^2}}{a_0}\right),\
%A_2 (r) = \frac{I_2(r)}{I_0(r)} = \left(\frac{\sqrt{a_2(r)^2 + b_2(r)^2}}{a_0}\right),\
%\frac{I_m(r)}{I_0(r)} =  \left(\frac{\sqrt{a_m(r)^2 + b_m(r)^2}}{a_0} \right),\
I_m(r) = \sqrt{a_m(r)^2 + b_m(r)^2}
\end{equation}
and the relative Fourier amplitude of the m$=$2 component is 
\begin{equation}
A_2(r) = \frac{I_2(r)}{I_0(r)}.
\end{equation}
We plot the profile of $A_2(r)$ in Figure~\ref{fig:radial_mod_strength}(b).
We define the bar strength $A_2$ as follows,
\begin{equation}
%A_2 \equiv \rm{max} \left( \frac{I_2(r)}{I_0(r)} \right).
A_2 \equiv \rm{max} \left( A_2(r) \right).
\end{equation}

\subsection{Identifying B/P bulges}
\label{sec:anal_bp}

\begin{figure*}[ht!]
\includegraphics[width=\textwidth]{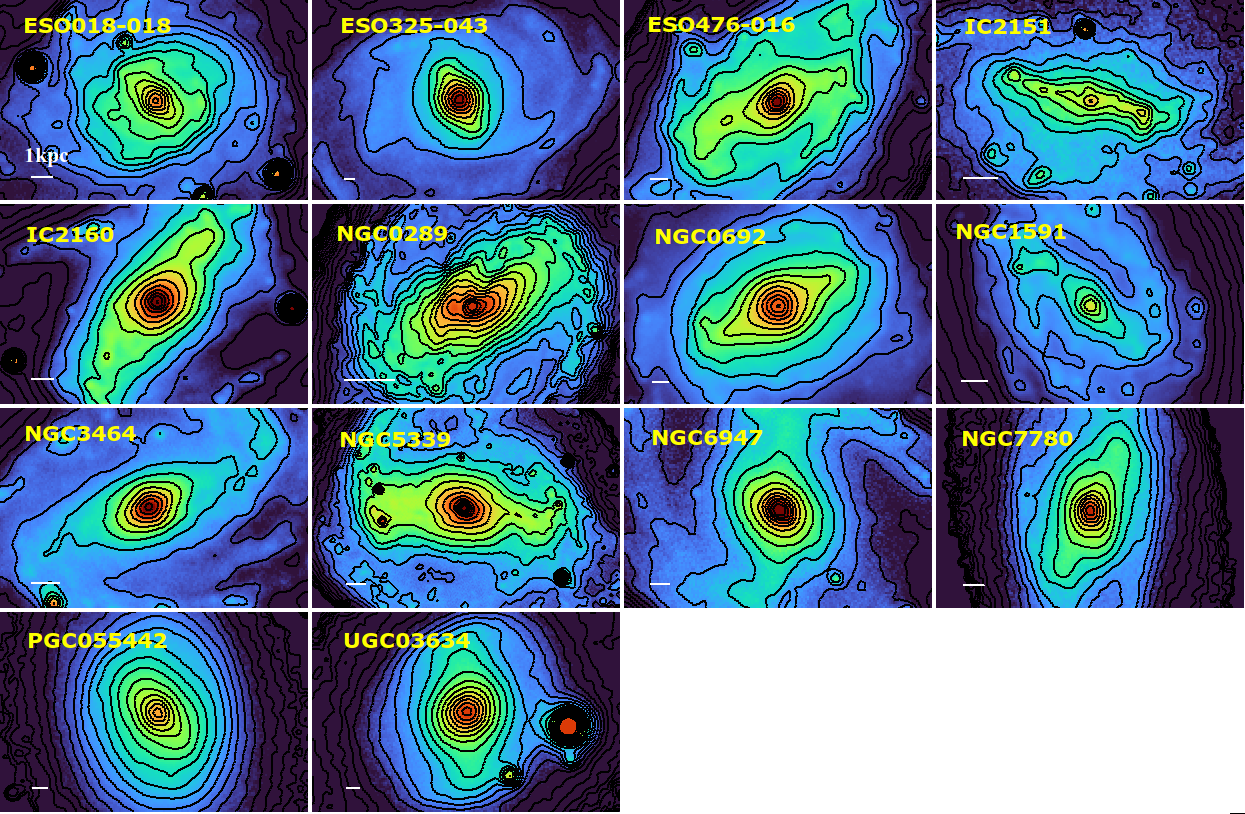}
\caption{Isophotes of galaxies in the bar region from the DESI and Pan-STARRS r-band images for UGC03634. The white line on the bottom left spans 1 kpc for each galaxy. Images are not corrected for deprojection, as this may distort the shape of the bulge.
\label{fig:contour}
}
\end{figure*}

Ideally, B/P features can be best traced in infrared images because prominent dust lanes in optical images can obscure the shape of bulge isophotes. For this reason, we utilize infrared images, if available, to determine the presence of a B/P feature. There are 5 objects available from the IRAC Camera channel 1 onboard the Spitzer Space Telescope, which has a typical FWHM of $1\farcs7$ -- $2\farcs0$ (\citealt{sheth_12, salo_15}) : NGC 0289 and NGC 5339 from the Spitzer Survey of Stellar Structure in Galaxies (\citealt{sheth_12}), ESO018-018, NGC 0692, and UGC03634 from the Spitzer Heritage Archive (PID: 61003).
%We also examined infrared $K_s$ band images from the 2MASS (\langle \rm FWHM \rangle $=$ 2.5 $\sim$ 3.0", \cite{skrutskie_06}).
However, due to the limited resolution or depth of the infrared images,  
we first examine the presence of B/P bulges using images from the DESI (Pan-STARR for UGC03634) which has a typical FWHM of $\sim$ 1.2" in the r-band (\citealt{dey_19}). We then double check the isophotes from the Spitzer, if available.

\begin{figure*}[htb]
\includegraphics[width=\textwidth]{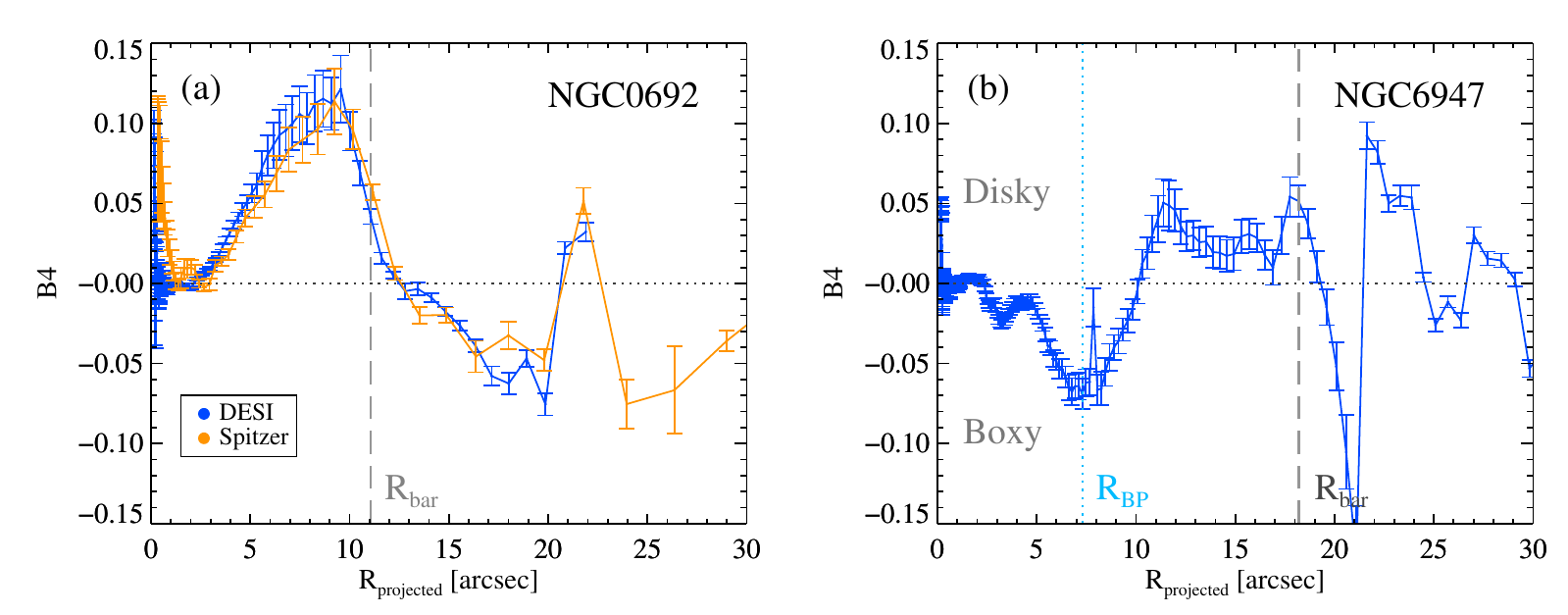}
\caption{Profiles of the boxiness parameter, B4, for two example galaxies having (a): a disky bulge and (b): a boxy/peanut (B/P) bulge. If B4 is positive, the isophote at that radius is disky. Conversely, if B4 is negative, the isophote is boxy. The blue points are from the DESI r-band images and orange points are from the Spitzer 3.6 $\mu$m image. Vertical dashed lines represents the bar radius ($R_{\rm bar}$) from the maximum ellipticity. The vertical blue dotted line indicates the size of the B/P bulge ($R_{\rm BP}$), measured from the minimum of B4 inside the bar radius. B4 profiles are measured on projected images to prevent artificial elongation of bulges due to the deprojection process. 
B4 profiles for the other samples are presented in the Appendix.
\label{fig:b4_profile}
}
\end{figure*}

In order to classify whether the galaxy has a B/P bulge, we first visually examine the images and isophotes of galaxies and later confirm with the quantity that describes the boxiness of the isophotes. Visual inspections are performed by two authors (TK and DAG).
We present the isophotal contours of the sample galaxies in Figure~\ref{fig:contour}. 
B/P bulges are puffed up perpendicular to the disk plane and thus are 3D structures and as such, the deprojection process may distort the shape of the bulge. Therefore, we do not deproject images to assess the presence B/P bulges.
In general, the isophotes of galaxies are not perfect ellipses.
If we model the isophotes of galaxies with the Fourier analysis, the higher order term (e.g., m$=$4 of Equation \ref{eg:fourier_eq2}) provides information about the shape of isophotes. 

To characterize the shape of isophotes, we employ \texttt{ellipse} task of the \texttt{STSDAS} package in the \texttt{IRAF}. The \texttt{ellipse} task performs isophotal analysis on galaxy images based on the analysis of  \citet{jedrzejewski_87}. 
%We follow the convention of the definition of Fourier analysis in \texttt{ellipse} task. Contour points of each isophotes C can be modeled as a function of azimuthal angle ($\phi$) as,

%\begin{equation}
%\begin{aligned}
%An = \frac{1}{\pi} \int C(\phi) \: sin (n\phi) \: d\phi
%\\
%Bn = \frac{1}{\pi} \int C(\phi) \: cos (n\phi) \: d\phi.
%\end{aligned}
%\end{equation}

In particular, if the term of $cos \, 4\theta$, which is usually named as ``B4'', is negative, the isophote appears to be boxy. On the contrary, if ``B4'' is positive, the isophote is disky. Detailed descriptions on obtaining the B4 parameter can be found in \citet{jedrzejewski_87} and \citet{erwin_13}. 

We present B4 profiles for two example galaxies, a galaxy without and with a B/P bulge (i.e.,``disky" and ``boxy") in Figure~\ref{fig:b4_profile}. We classify galaxies to have a B/P bulge if the B4 is negative inside the bar region for at least longer than a quarter of the bar length.  
B4 profiles for the whole samples are available in the Appendix.
In Figure \ref{fig:b4_profile}, NGC 0692 shows positive B4 inside the bar radius. Thus, we classify the galaxy to have a disky bulge. On the other hand, NGC 6947 shows negative B4 inside the bar radius and thus we classify the galaxy to have a B/P bulge. In our sample, IC 2151 does not have a prominent bulge, thus we exclude this galaxy from the B/P group even if it shows negative B4 within the bar region. In NGC 0289, dust lanes are prominent in the optical images which may distort the isophotes of galaxies. Thus, we double check with infrared images from the Spitzer, if available. We present B4 profiles of all the sample galaxies in the Appendix.

We found 6 galaxies with a B/P signature out of 14 sample galaxies. The presence of a B/P bulge is tabulated in Table 1. 
The fraction of B/P bulges in our sample is 43\%, which somewhat smaller than the findings of \citet{erwin_13}, who find at least two-thirds of S0–Sb bars exhibit buckled or B/P structures. \citet{erwin_13} also note that the detection rate of B/P bulges is higher in highly inclined galaxies and when the bar is aligned closer to the major axis. Consequently, our ability to detect B/P bulges may be limited, as our sample consists of moderately inclined galaxies. Furthermore, our sample is not statistically complete, which could result in a different B/P bulge fraction.

%<B4 구하는 방법 설명 추가 하기>
%\begin{equation}
%    I(R, \theta)= I_0 + \sum_{n=1}^{\infinite}{m_i} \sin(m\theta_{i}),
%\end{equation}

%\begin{figure*}[ht!]
%\includegraphics[width=\textwidth]{rst_comp_set3.eps}
%\caption{The same figure as Fig.\ref{fig:nc_qb} but with presence of a  boxy/peanut (B/P) bulge. Galaxies with a strong B/P bulge are plotted in dark blue squares, and galaxies with a mild B/P bulge are plotted in sky blue squares. Two light gray points represent galaxies that has the bar aligned with the major/minor axis of the disk. 
%\label{fig:nc_qb2}
%}
%\end{figure*}

\section{Non-circular motions and the bar strength} \label{sec:nc_barstrength}
\begin{figure*}[ht!]
\includegraphics[width=\textwidth]{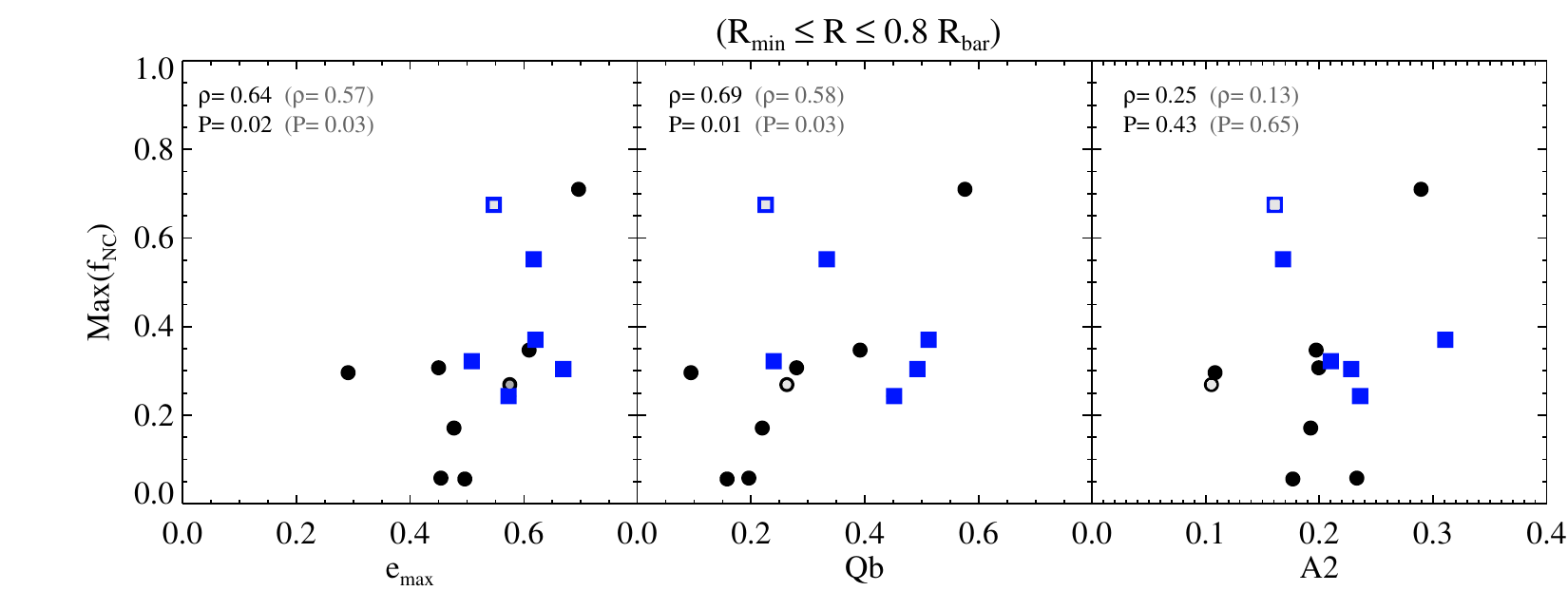}
\includegraphics[width=\textwidth]{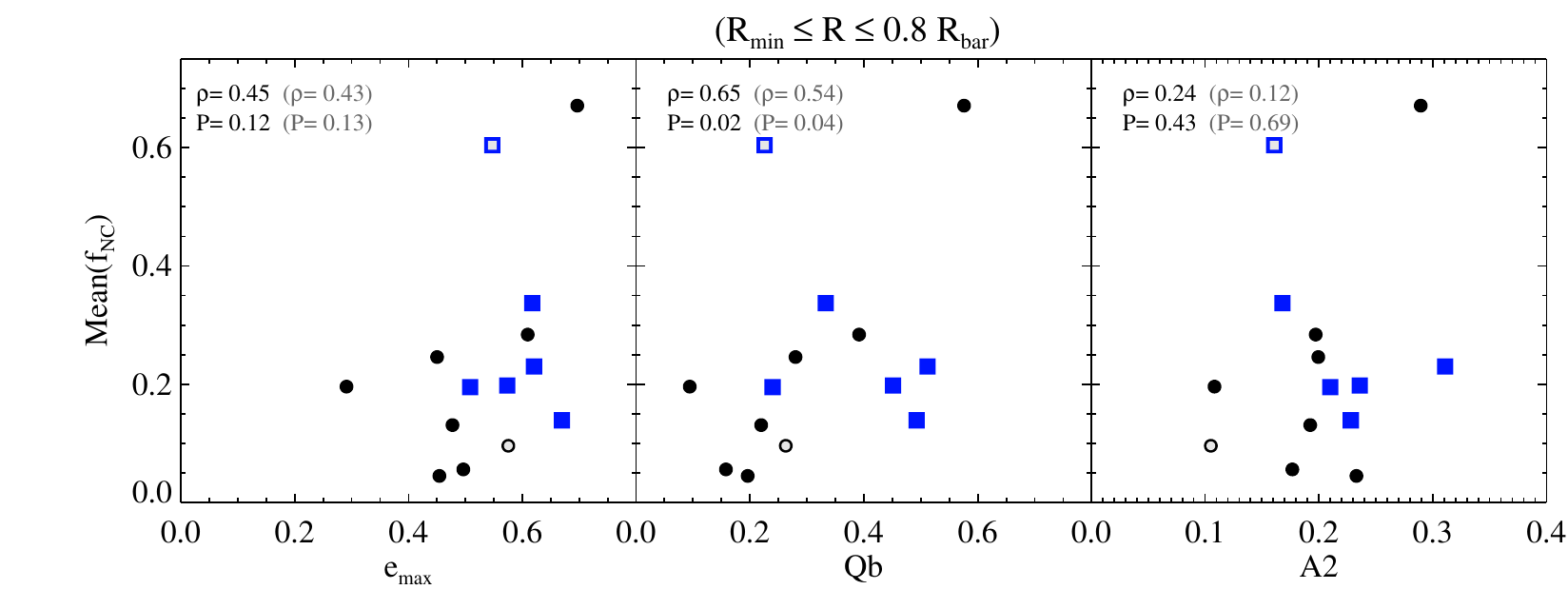}
\includegraphics[width=\textwidth]{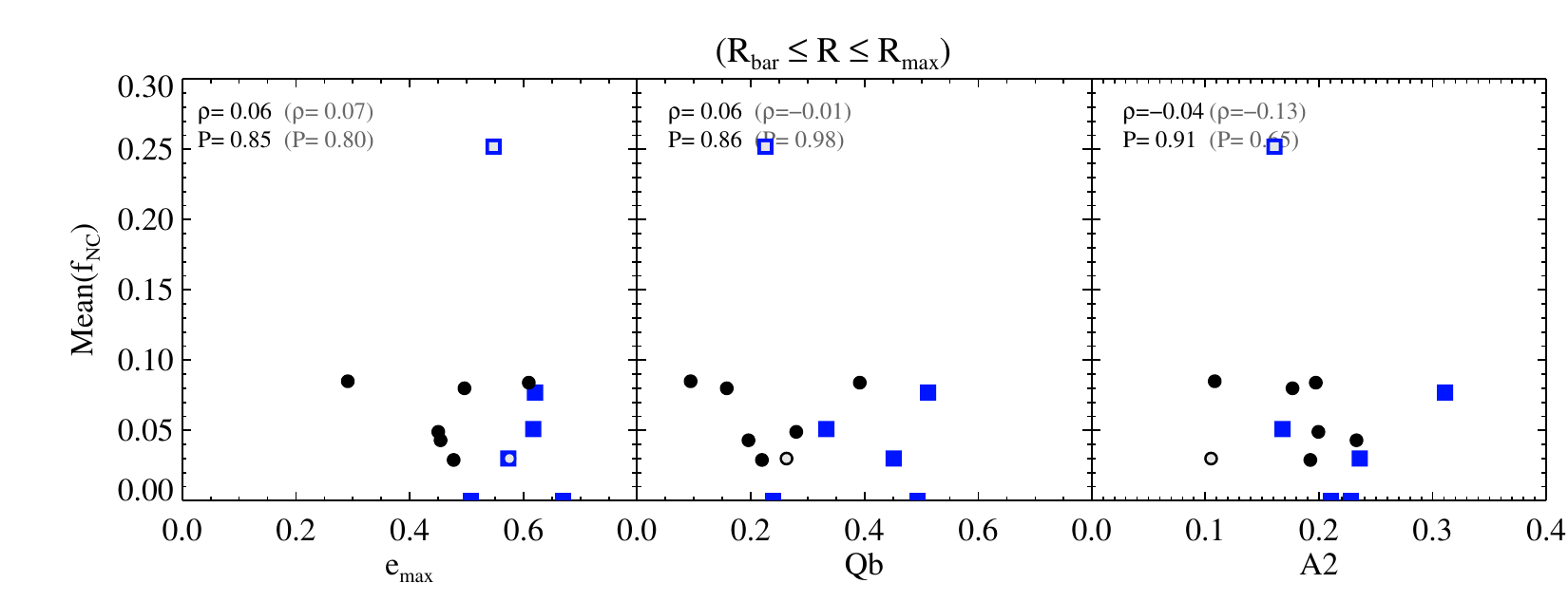}
\caption{The comparisons of bar strength (bar ellipticity, \qb~and $A_{2}$ ) and the degree of non-circular motions.
We present the maximum and the mean ratio of the non-circular velocity over circular velocity, Max($f_{\rm NC}$) and Mean($f_{\rm NC}$), respectively, which are indicators of the degree of non-circular motions. 
$R_{\rm min}$ is the larger one between 2" (the inner most annulus of the bisymmetric fit) and 0.2$R_{\rm bar}$ to avoid nuclear structures. 
$R_{\rm max}$ is set to the radius where both $V_{\rm rad}$ and $V_{\rm tan}$ become zero for more than two consecutive annuli, which may go beyond the bar radius.
Blue squares denote galaxies with a B/P bulge. 
%Galaxies are plotted in light grey if the bar is aligned with the major axis of the galaxy ($\rm PA(disk) - \rm PA(bar) < 10^{\circ}$ or $> 80^{\circ}$ ).
Galaxies are plotted in light grey (one square and one circle) if the alignment between the bar and the major axis of the galaxy is within 10 degrees, since in these cases the determination of $f_{\rm NC}$ is compromised (see text).
The Spearman correlation coefficients are shown in each panel. 
The Spearman correlation coefficients in black represent samples whose bars are not aligned with the major axis of the galaxy whereas those in grey represent include all galaxies. Note that the range of $A_2$ is different from that of $e_{\rm max}$ and \qb.
\label{fig:nc_qb}
}
\end{figure*}

We examine whether there is a relation between the bar strength and the degree of non-circular motions of barred galaxies. In Figure~\ref{fig:radial_mod_strength}(b) and Figure~\ref{apfig:rad_profile}, we present the radial profiles of the fraction of the non-circular motions to circular rotation, $f_{\rm NC}$ which is defined in Equation \ref{eq:fnc}.
Within our sample, the peak of $f_{\rm NC}$ occurs on average at 0.45$R_{\rm bar}$ with a standard deviation of 0.28. $\rm R_{\rm bar}$ denotes the bar radius derived from the maximum of radial ellipticity profiles on deprojected images. This finding suggests that, compared to circular rotations, non-circular motions are strongest slightly below half of the bar radius, on average.

There are no prior studies on stellar kinematics that can be directly compared with our results. \citet{salak_19}, using CO (J=1-0) velocity maps and Fourier analysis for 7 barred galaxies, find that the maximum of non-circular motions typically occurs at a radius of 0.3$R_{\rm bar,v}$, on average, where $R_{\rm bar,v}$ represents the visually determined bar radius. This maximum occurs at a slightly shorter radius compared to our findings. Several factors may contribute to these differences. First, non-circular motions in gas and stars may differ. Second, the methods used to assess non-circular motions vary, with our study using a bisymmetric model, whereas \citet{salak_19} employed Fourier analysis. Third, the methods for estimating bar length differ: we use ellipticity profiles to estimate the bar length, while \citet{salak_19} adopted visual estimation. \citet{erwin_05_bar} finds that the bar length derived from the maximum ellipticity, which we used to determine our bar lengths, can underestimate the true length of the bar (see also \citealt{wozniak_95, athanassoula_02a, laurikainen_02_2mass,erwin_03, aguerri_09}). Therefore, these effects might explain the differences between our results and those from \citet{salak_19}.

%Using CO (J=1-0) velocity maps for 7 barred galaxies with Fourier analysis, \citet{salak_19} find that the maximum of non-circular motions occurs at the radius of 0.3$R_{\rm bar,v}$ on average where $R_{\rm bar,v}$ is visually determined bar radius. They found that the maximum of the non-circular motions occurs at slightly shorter radius compared to our data. There are three elements which could lead slightly different results. First, non-circular motions of gas and stars can be different. Second, the method to assess non-circular motions are different, the bisymmetric model and the Fourier analysis. Third, the method of estimating bar length are also different. \citet{erwin_05_bar} finds that the bar length derived from the maximum ellipticity can underestimate the true length of the bar (see also \citealt{wozniak_95, athanassoula_02a, laurikainen_02_2mass,erwin_03}). Therefore, these effects might explain the the differences between our results and those from \citet{salak_19}.

We find that the radial motions ($V_{\rm 2,r}$) are comparable to, or slightly greater than, the tangential motions ($V_{\rm 2,t}$) as can be seen from Figure~\ref{fig:radial_mod_strength}(b) and Figure~\ref{apfig:rad_profile}. This is consistent with results from pseudo observations of galaxies in numerical simulations (\citealt{randriamampandry_15,randriamampandry_16}).

We take the maximum and the mean of $f_{\rm NC}$ (Max($f_{\rm NC}$) and Mean($f_{\rm NC}$)) for each galaxy as an indicator of the degree of non-circular motions driven by bars and compare them with the various bar strength indicators in Figure~\ref{fig:nc_qb}.
In estimating the Max($f_{\rm NC}$) and Mean($f_{\rm NC}$), we exclude central regions (the larger of either the inner 0.2$R_{\rm bar}$ or r $\leq$ 2", which is the innermost annulus of the bisymmetric fit) as there can be an unresolved nuclear disk, nuclear spirals, or even nuclear bar. 
%We set two different outer boundaries to estimate Max($f_{\rm NC}$) and Mean($f_{\rm NC}$). First, 
In estimating Max($f_{\rm NC}$) and Mean($f_{\rm NC}$), we set the outer boundary to 0.8$R_{\rm bar}$ in order to avoid outer ring or spiral arms if present. We present the relation between Max($f_{\rm NC}$) and the bar strength in the first row of Figure~\ref{fig:nc_qb}. 

% We present the relation between Max($f_{\rm NC}$) and the bar strength in the first row of Figure~\ref{fig:nc_qb}. 
%The Max($f_{\rm NC}$) for each galaxy remains unchanged when applying the two different outer boundaries, either 0.8$R_{\rm bar}$ or $R_{\rm max}$, except for IC2151 and ESO018-018. For these two galaxies, Max($f_{\rm NC}$) increases by $\sim$0.1 when the outer boundary is extended from 0.8$R_{\rm bar}$ to $R_{\rm max}$. Therefore, we present Max($f_{\rm NC}$) within 0.8$R_{\rm bar}$ only.
%We \textbf{plot} the Mean($f_{\rm NC}$) with two different outer boundaries in the second and third row of Figure~\ref{fig:nc_qb}. The Spearman's rank correlation coefficients ($\rho$) and the statistical significance (P) are presented in the upper left corner of each panel.
We plot the Mean($f_{\rm NC}$) with the outer boundary at 0.8${R_{\rm bar}}$ in the second row of Figure~\ref{fig:nc_qb}. 
The Spearman's rank correlation coefficients ($\rho$) and the statistical significance (P) are presented in the upper left corner of each panel.

We find a strong correlation between the bar strength (bar ellipticity and $Q_b$) and the maximum and the mean of $f_{\rm NC}$ out to $0.8R_{\rm bar}$. This implies that galaxies with stronger bars exhibit more enhanced non-circular motions. 
To our knowledge, this is the first observational study to find the link between the bar strength and the amplitude of stellar non-circular motions driven by bars.

We also define another outer boundary $R_{\rm max}$ where both $V_{\rm rad}$ and $V_{\rm tan}$ velocity become zero for more than two consecutive annuli along the radius. As shown in Figure~\ref{fig:radial_mod_strength}(a) and Figure~\ref{apfig:rad_profile} in the Appendix, $V_{\rm rad}$ and $V_{\rm tan}$ are still non-zero after the bar radius. This might be due to the non-circular motions from spiral arms. Thus, the $R_{\rm max}$ may extend beyond the $R_{\rm bar}$ and may encompass the outer ring or spiral arms if present. However, we note that $R_{\rm max}$ occurs at a smaller radius than $R_{\rm bar}$ for NGC 6947 and NGC 7780.

%In the third row of Figure~\ref{fig:nc_qb}, we find an even stronger relation (higher $\rho$) between the bar strength and the degree of non-circular motions. This might indicate that stronger bars are more closely link to spiral arms, but further studies should be carried out to draw a clear conclusion. The correlation between the $f_{\rm NC}$ and $A_2$ is found to be weak compared to the bar ellipticity and \qb. This might be due to the narrow range of $A_{2}$, and the sensitivity to bulge properties and spiral arms of $A_{2}$ parameter.

%As the third row of Figure~\ref{fig:nc_qb} shows, we find a slightly stronger relation (higher $\rho$) between the bar strength and the degree of non-circular motions when the outer boundary for calculating Mean($f_{\rm NC}$) is extended to $R_{\rm max}$. This might indicate that stronger bars tend to be longer and have their maximum \qb values occurring at larger r, thereby providing stronger perturbations for stars even in regions outside the bars. Further studies are necessary to draw a clear conclusion on this issue. 

As Figure ~\ref{fig:radial_mod_strength}(a) shows, the non-circular motions are much stronger within the bar radius compared to the outer regions. To examine the non-circular motions in the outer region, we estimate the Mean($\rm{f_{NC}}$) between $R_{\rm bar}$ and $R_{\rm max}$, and plot them as a function of bar strength in the bottom panel of Figure~\ref{fig:nc_qb}. Interestingly, there is no clear relation between the bar strength and the level of non-circular motions in $R_{\rm bar} < R < R_{\rm max}$. 
Beyond the bar, we find that non-circular motions are weak at the level of $f_{\rm NC}$ $\lesssim$ 0.1. This implies that non-circular motions produced by spiral arms are relatively weak.

It is unclear why the correlation between $f_{\rm NC}$ and $A_2$ is found to be weak compared to the bar ellipticity and \qb, but this might be due to narrow range of $A_2$ of our sample galaxies or $A_2$ might be more sensitive to bulge (figure 12 of \citealt{diaz-garcia_16a}) and/or spiral structures. 

%\subsection{Limits on the PA of the bar: Relative angle between the PA of the bar and major/minor axis of the disk}

The velocity map of a barred galaxy is a combination of circular motions originated from the galaxy rotation and non-circular motions due to the bar. \citet{randriamampandry_15} find that the tilted-ring method under/overestimates the circular motions when the bar lies parallel/perpendicular to the projected major axis of the galaxy (see also \citealt{oman_19}). 
They also show that even if we account for non-circular motions by adopting 
the bisymmetric model, it may produce large rotation velocity with a large error if the bar is aligned with either the major or minor axis of the bar (See also \citealt{sellwood_10}). In two of our sample galaxies, the bar lies close to the major axis of the galaxy (NGC 0289 and NGC 3464), and we plot them in light grey in Figure~\ref{fig:nc_qb}. The Spearman's rank correlation coefficients including these two points are shown in light grey in parenthesis. 

A previous study on the relation between bar strength and non-circular motions of barred galaxies has reached somewhat different results.
Using H$\alpha$ velocity map, \citet{erroz-ferrer_15} modeled circular velocity of galaxies. They subtract the circular velocity model map from the observed velocity field. They obtained the degree of non-circular motions by analyzing the cumulative distribution functions of the residual velocity map. They find that the amplitude of non-circular motions do not correlate with bar strength (\qb) which is different from our results. However, there are two important differences to consider. First, we examined the non-circular motions from stellar velocity maps. Deriving kinematic parameters such as circular, radial, tangential velocities based on H$\alpha$ velocity maps might be hampered because ionized gas suffers effects from forces other than gravity (e.g., stellar winds, supernova feedback, shocks).
Second, we derive bar-driven non-circular motions by fitting the bisymetric model.
Thus, these two differences might have lead to different results.

Stars in a barred galaxy are subject to the gravitational potential of the bar.  In particular, stars trapped in the bar follow non-circular orbits. Initially, these non-circular motions were small, but they gradually intensified, strengthening and elongating the bar, which in turn further amplified the non-circular motions. This feedback mechanism ultimately leads to a quasi-equilibrium state, which we observe in the present. In this paper, we aim to verify that strong bars exhibit strong stellar non-circular motions which also include non-circular motions in the bar itself, rather than focusing on the influence of the bar on stars within the bar radius but outside the bar structure. Our methodology, which employs a light-weighted average velocity map at each radius, is unable to distinguish between stars moving along the bar and those in circular orbits within the inner disc inside the bar radius. Instead it models velocity components, whether they are associated with circular or non-circular motions. Although a strong bar may dominate in the region within the bar radius, there will be stars still in circular orbits within the bar radius which have not been trapped by the bar yet. In regard to this, to explore the kinematic differences between stars moving along the bar structure and those moving in circular orbits in the inner disk inside the bar, \citet{walo-martin_22} have demonstrated that bars induce significant differences of velocity dispersion along the bar major and minor axes with cosmological zoom-in simulations. Although this analysis would be feasible for line-of-sight velocity dispersion with our observational dataset, it lies beyond the scope of our current study and could be addressed in a future study.

Our results are consistent with cosmological zoom-in simulations (\citealt{Pinna_18, walo-martin_22}), which show, from an overall perspective, that stronger bars lead to more pronounced non-axisymmetry in stellar kinematics.

\section{Role of boxy/peanut bulge in non-circular motions} \label{sec:bp} 
\begin{figure*}[ht!]
\includegraphics[width=\textwidth]{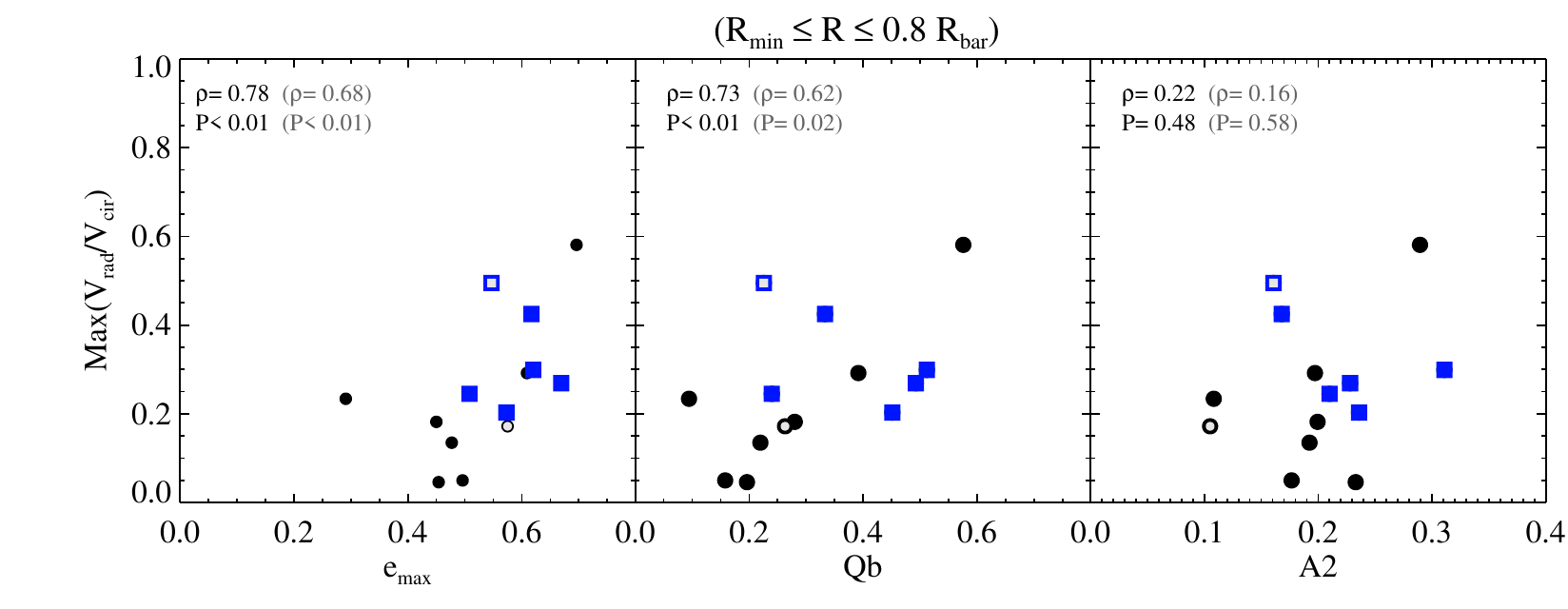}
\includegraphics[width=\textwidth]{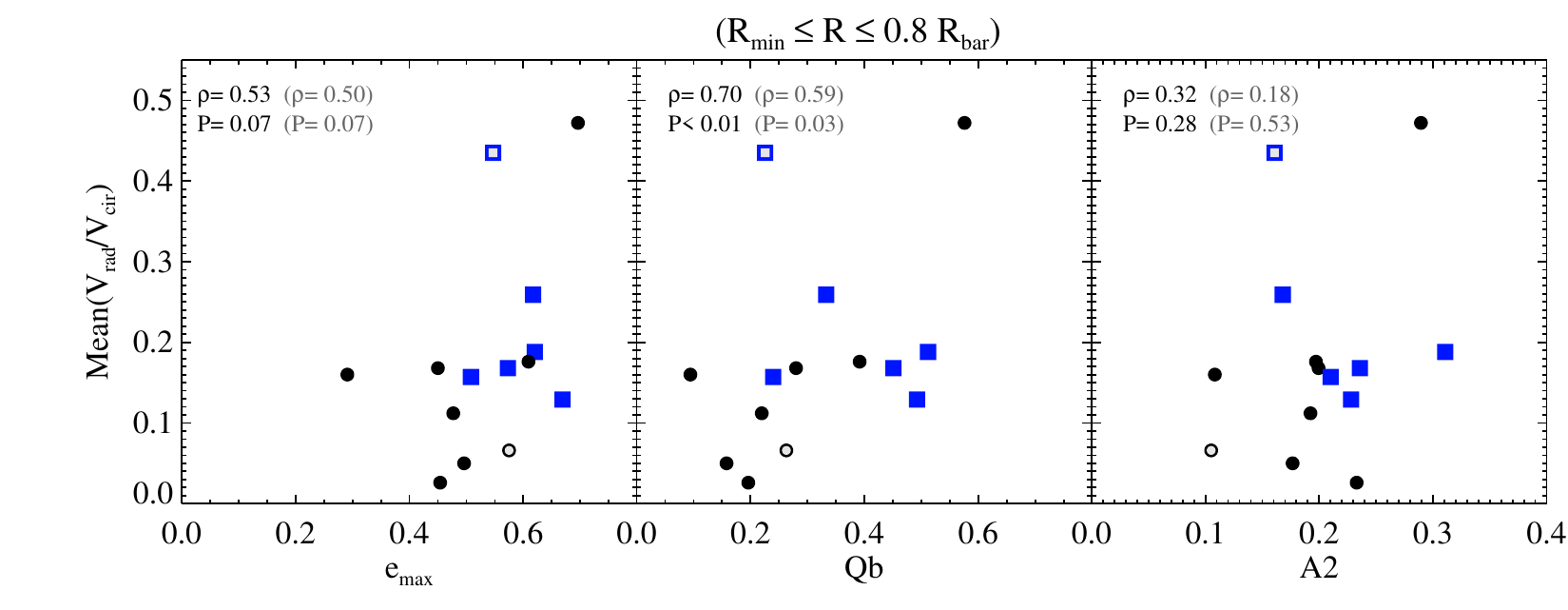}
\includegraphics[width=\textwidth]{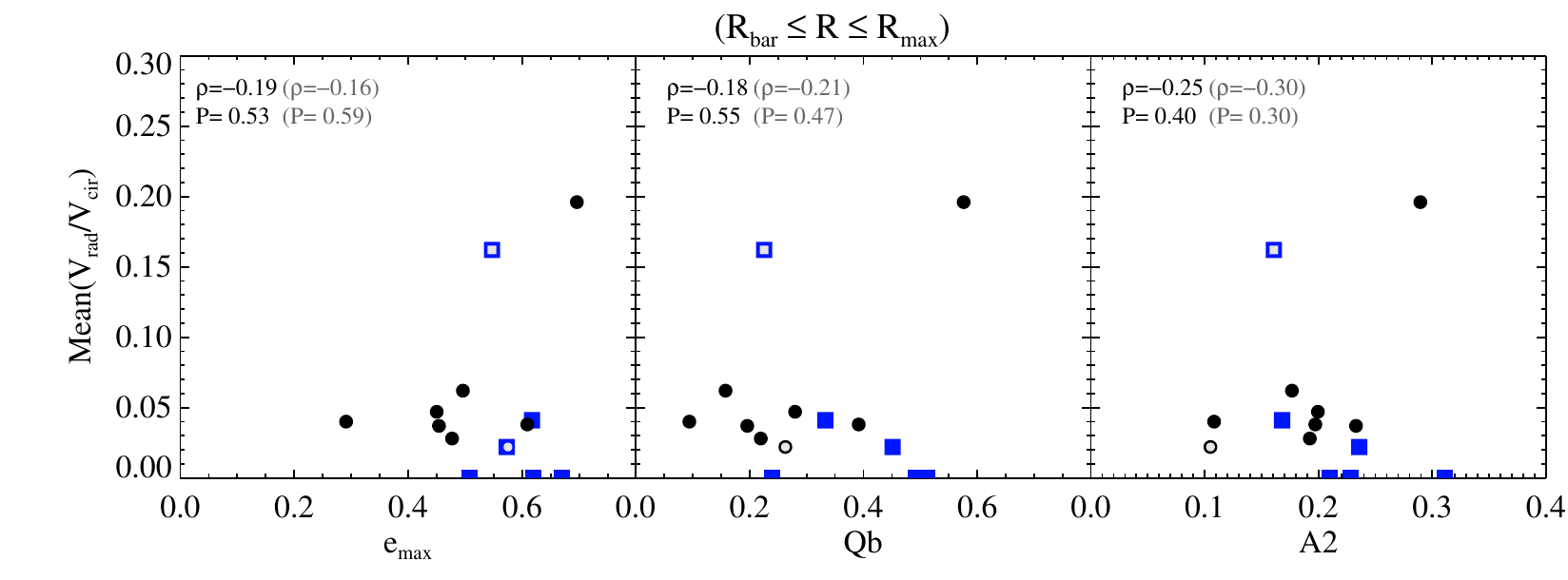}
\caption{The same as Figure~\ref{fig:nc_qb}, but now bar strengths and the maximum and the mean of stellar radial velocity ($V_{\rm rad}$) over circular velocity ($V_{\rm cir}$). 
%Galaxies with a B/P features are plotted in blue squares. Two light gray points (one square and one circle) represent galaxies whose alignment between the bar and the major axis of the galaxy is within 10 degrees.
\label{fig:nc_qb2}
}
\end{figure*}

We explore the impact of having a B/P bulge on the degree of non-circular motions and the location of maximum non-circular motions in this section. 
We plot the mean stellar radial velocity over the circular velocity, Mean($V_{\rm rad}/V_{\rm cir}$), in Figure~\ref{fig:nc_qb2}, where $V_{\rm rad}$ is $V_{\rm 2,r}$ and $V_{\rm cir}$ is $V_{\rm t}$ from Equation \ref{eq:bis}.
 First, the figure shows that there is a strong relation between the bar strength and the mean radial velocity to circular velocity ratio.
This indicates that stronger bars produce stronger stellar radial motions which is consistent with the results we obtain from the previous section.
Second, we examine the impact of the presence of a B/P bulge on the radial motions. We plot galaxies with a B/P bulge in blue squares.
We find that galaxies with a B/P bulge show higher Mean($V_{\rm rad}/V_{\rm cir}$) than galaxies without a B/P bulge do as a group by $\sim45\%$, and higher Max($V_{\rm rad}/V_{\rm cir}$) by $\sim50\%$.
Also Mean($f_{\rm NC}$) of galaxies with a B/P bulge are found to be higher by $\sim30\%$ than that of galaxies without a B/P bulges are, and Max($f_{\rm NC}$) of galaxies with a B/P bulge are higher by $\sim50\%$ than that of galaxies without a B/P bulge.

%Also galaxies with a stronger B/P have slightly higher Mean($V_{\rm rad}/V_{\rm cir}$) than galaxies with a mild B/P do on average.
%Stars form the backbone of the bar and also the B/P bulge, and their stellar orbits have been explained with various orbital families (e.g., \citealt{contopoulos_89b, patsis_02, wozniak_09, fragkoudi_15}). Therefore, our results indicate that orbital families that build B/P bulges may promote enhanced stellar radial motions in barred galaxies.

We also investigate the relation between the location of maximum of the non-circular to circular motions and the size of B/P bulges. For galaxies with a B/P bulge, we define a characteristic size of a B/P bulge ($R_{\rm BP}$) along the major axis of the galaxy where the radial B4 parameters from the \texttt{IRAF ellipse} task reaches its minimum inside the bar radius. We denote $R_{\rm BP}$ in blue dotted line in Figure~\ref{fig:b4_profile}. The radius with the minimum B4 means that the isophote at the radius exhibits the most boxy feature. We find that, on average, B/P bulges span $0.40\pm 0.12 \times R_{\rm bar}$. This is consistent with the results of \citet{erwin_13} who find that B/P bulges span 0.38 times the bar radius. 

Non-circular motions ($V_t,V_{\rm 2,t},V_{\rm 2,r}$, Max($f_{\rm NC}$) and Mean($f_{\rm NC}$)) and bar strengths (bar ellipticity, \qb, and $A_2$) are all measured on deprojected images. However, for the analysis of B/P structures, deprojecting images can cause artificial elongation of bulge components, which may interfere with accurately measuring the radius of the B/P structures. Therefore, we first measure $R_{\rm BP}$ on projected images and then deproject $R_{\rm BP}$ analytically following the Equation 1 of \citet{gadotti_07}.

We find that for galaxies with a B/P bulge, the maximum ratio of non-circular to circular velocity, Max($f_{\rm NC}$), occurs at the radius of 0.37$\pm0.20\times R_{\rm bar}$, which is slightly lower compared to those of the galaxies without a B/P bulge (0.52$\pm0.32\times R_{\rm bar}$). Thus, we find that the maximum of non-circular to circular velocity occurs slightly shorter for galaxies with a B/P bulge, but this has to be confirmed with larger samples.
%For galaxies with a B/P bulge, we deproject the B/P radius analytically, following the Equation 1 of \citet{gadotti_07}. Then, 
We find that Max($f_{\rm NC}$) occurs at 1.02$\pm 0.53 R_{\rm BP}$. This may imply that the size of B/P bulges could be related to the position of the maximum of non-circular motions.

%% NOTE: f_NC, QB,A2, ellipticity are obtained on deprojected, but BP properties are obtained on projected images to avoid elongation of the B/P bulges! 

Stars form the backbone of the bar and also the B/P bulge, and their stellar orbits have been explained with various orbital families (e.g., \citealt{contopoulos_89b, patsis_02, quillen_02, wozniak_09, fragkoudi_15}). Therefore, our results, which show a higher degree of non-circular motions in galaxies with a B/P feature and indicate Max($f_{\rm NC}$) occurs near the B/P radius, may suggest B/P bulges promote enhanced stellar radial motions in barred galaxies. However, we are rather cautious about this interpretation and it warrants further scrutiny. In our sample, the majority of galaxies with a B/P feature have strong bars. Therefore, there is a high chance that the galaxies with a B/P structure exhibit stronger non-circular motions simply because they have strong bars.
This hypothesis can be tested through numerical simulations with various setups and also through further observational studies on a sufficiently large number of galaxies covering various bar strengths with and without B/P features for more statistically robust comparisons.

There are not many studies conducted on the impact of the presence of B/P bulges on the non-circular motions of barred galaxies.
\citet{fragkoudi_16} find that the presence of B/P bulges can reduce the rate of gas inflow by an order of magnitude. The impact of B/P bulges on gas inflow can be best traced by gaseous kinematics. However, the limited S/N of H$\alpha$ in our datasets for the bar region do not allow us to obtain reliable modeling of the gaseous bisymmetric motions, and we are only able to get stellar kinematics. Thus, we cannot directly compare our results from stellar kinematics with previous studies.
%Our findings from stellar kinematics point towards different direction compared to the results for gas kinematics from \citet{fragkoudi_16}. But we stress that results from stellar and gas kinematics could be different. 
%Non-circular motions from gaseous kinematics are found to be more pronounced compared to stellar kinematics (\citealt{lopez-coba_22}).
Therefore, further studies with more statistically complete samples using both stellar and gaseous kinematics are necessary to solidify our understanding of the impact of having a B/P bulge on the non-circular motions of galaxies.

\section{What cause the scatter in the bar strength -- non-circular motions?}
We find a strong relation between the bar strength and the degree of non-circular motions in Figure~\ref{fig:nc_qb}, also between the bar strength and the stellar radial velocity to circular velocity ratio in Figure~\ref{fig:nc_qb2}.
However, there is some scatter around these relations. In this section, we examine the possible factors that cause the scatter between the bar strength and the degree of non-circular motions.

%The bisymmetric model is constructed to fit non-circular motions driven by bar-like or oval distortions to an axisymmetric potential, and the bisymmetric model has been used to suitably describe the velocity map of barred galaxies (\citealt{spekkens_07, sellwood_10, holmes_15, lopez-coba_22}). However, as gas flows inwards mainly along the bar dust lane (e.g., \citealt{athanassoula_92b, englmaier_97, kim_w_12a, sormani_15b, seo_19, sormani_23}), the bisymmetric model may not perfectly account for this inflow concentrated along the bar dust lane, especially for the gaseous component (See also \citet{lopez-coba_24} for an inflow model).

There are certain assumptions we made in using the bisymmetric model that may not accurately represent the kinematics of barred galaxies.
The bisymmetric model neglects higher harmonics, and there may be effects from these higher harmonics. However, as the m$=$2 term is the largest one (e.g., \citealt{spekkens_07}), the effect of higher order terms would be limited.
We also assume disks to be flat, having the same inclination at all radii, which may not necessarily be the case for galaxies that have warps (\citealt{reshetnikov_02, sanchez-saavedra_03, ann_06, vanderkruit_11}). Our samples do not show obvious warps, but as our samples are at moderate inclinations there might be warps to some degree that may affect the observed velocity fields. However, the impact of warps would be confined to the outer regions of galaxies.

%Furthermore, the bisymmetric model assumes the disk to be flat, having the same inclination at all radius, which may not be necessarily the case for galaxies which have warps (\citealt{reshetnikov_02, sanchez-saavedra_03, ann_06, vanderkruit_11}). Our samples do not show obvious warps, but as our samples are at moderate inclinations there might be warps to some degree which may affect the observed velocity fields.

The bisymmetric model is designed to fit non-circular motions driven by bar-like or oval distortions within an axisymmetric potential, and has been used to suitably describe the velocity map of barred galaxies (\citealt{spekkens_07, sellwood_10, holmes_15, lopez-coba_22}). However, it is necessary to evaluate how well the bisymmetric model can reproduce the stellar non-circular motions for various types of barred galaxies using numerical simulations. It should also be noted that gas inflow predominantly occurs along the bar dust lane (e.g., \citealt{athanassoula_92b, englmaier_97, kim_w_12a, sormani_15b, seo_19, sormani_23}), which the bisymmetric model may not fully capture, particularly for the gaseous component (see details for an inflow model in \citealt{lopez-coba_24}).

Non-circular motions can be induced by various processes other than a bar, such as spiral arms, warps, tidal interactions, and a triaxial dark matter halo (\citealt{bosma_78, begeman_87, rhee_04, hayashi_06, trachternach_08, stark_18, oman_19}). Although the impact of these other processes may not be significant, especially given our selection of barred galaxies without clear signs of interaction or warp, there may still be some slight effect on the magnitude of non-circular motions within the bar region. Consequently, while these factors are likely negligible, they could contribute minimally to the scatter in the relation between the bar strength and the degree of non-circular motions.

In constructing the gravitational potential map to estimate \qb, we assumed that the M/L is constant which has been applied to many studies (e.g., \citealt{quillen_94,buta_01,salo_99,laurikainen_02_2mass, zhang_07, diaz-garcia_16a, diaz-garcia_16b, lee_20}). While M/L can vary with radius due to the dark matter halo, stellar populations (age, metallicity, star formation history) gas, and dust distribution (e.g.,\citealt{persic_96, bell_00, helfer_03, deblok_08, walter_08, munoz_mateos_09, bershady_11, kennicutt_12, querejeta_15, garcia-benito_19}), these variations are not expected to significantly impact our results. Indeed, \citet{diaz-garcia_16a} have shown that the effects of dark matter halos are not significant for galaxies with earlier Hubble types (T$\leq$5), which constitute the majority of our sample. They also find the effects of non-stellar contaminants to be 10--15$\%$, which is not significant. Consequently, the assumption of a constant M/L provides a reasonable approximation for our analysis. Incorporating radial variations in M/L is beyond the scope of this paper, but future studies could explore this aspect in more detail.

%However, M/L varies with radius due to the effects from dark matter halo, stellar populations (age, metallicity, star formation history) gas, and dust distribution (e.g.,\citealt{persic_96, bell_00, helfer_03, deblok_08, walter_08, munoz_mateos_09, bershady_11, kennicutt_12, querejeta_15, garcia-benito_19}). This might bring changes in \qb~and thus errors in the explored relation between \qb~and the degree of non-circular motions. Indeed, \citet{diaz-garcia_16a} explores the effects of dark matter halos and find that these effects are not significant for galaxies with earlier Hubble types (T$\leq$5), which constitute the majority of our sample. They also find the effects of non-stellar contaminants to be 10--15$\%$, which is not significant.

%\citet{fragkoudi_15} find that the bar strength, in particular \qb~is reduced when there is a B/P bulge in the galaxy. The degree of reduced bar strength depends on the length, width and scale height of the B/P structure. Therefore, it is not trivial to estimate how much the bar strength may have been reduced for our sample. However, as our samples with a B/P bulge show relatively higher values of the Mean($V_{\rm rad}/V_{\rm cir}$), the relation between the bar strength and the degree of non-circular motions would still be maintained even if the corrections due to B/P bulges are applied, resulting in higher bar strength to some degree.

The bar strength is also influenced by the presence of a B/P bulge. \citet{fragkoudi_15} find that radial profiles of $Q_T$ are modified in the presence of a B/P feature, although the {\qb} parameter may not be sufficiently sensitive to capture these changes. Using an updated bar strength indicator, they found that the bar strength decreases when a B/P structure is present. The degree of reduced bar strength depends on the length, width and scale height of the B/P structure. Therefore, it is not trivial to estimate how much the bar strength may have been reduced for our sample. However, asthe galaxies in our sample with a B/P bulge show relatively higher values of the Mean($V_{\rm rad}/V_{\rm cir}$), the relation between the bar strength and the degree of non-circular motions would still be maintained even if the corrections due to B/P bulges are applied, resulting in higher bar strength to some degree.

Finally, the uncertainties in estimating bar strength (bar ellipticity, \qb and $A_2$) may introduce scatter. These uncertainties include assumptions on the ratio of M/L, the scale height and length of the disk, and the deprojection process. Additional factors such as the effects of dust extinction, bulge prominence, and strong spiral arms connected to the bar may also contribute to the scatter.

All of the aforementioned factors may cause scatter in the diagram relating bar strength and non-circular motions. Nevertheless, we still observe a strong relation between the bar strength and bar-driven non-circular motions, which implies that they are closely linked, as expected.

%\section{Shock and the bar strength}
%We measure H$\alpha$ velocity jump across the dust lane. Pseudo slits are put across the dust lane which is defined using optical images. If the dust lanes are not prominent, we put slits across the bar. We show pseudo slits on the optical r-band image and H$\alpha$ velocity map in Fig. XXX, and plot the velocity jumps in Fig. XXX.

%\section{Other possible factors}
%\begin{itemize}
%\item Amount of available gas to flow inward
%\item Impact of evolutionary stage of bar on gas inflow
%\item Bulge prominence: e.g., Error-Ferrer+15: bulgeless galaxies show %various value! Why?
%\end{itemize}
%\\

\section{Conclusion and Summary} \label{sec:summary}

We examine the impact of bar strength on the kinematics of barred galaxies. 
Specifically, we measure non-circular motions by applying the bisymmetric model to the stellar kinematics of 14 barred galaxies observed with the VLT/MUSE.
We measure bar strength using r-band images from the DESI/Pan-STARRS.
We present the first observational results that show bar strength is closely related to the degree of non-circular motions in barred galaxies. 
This implies that stronger bars induce stronger stellar non-circular motions. The ratio of non-circular to circular velocity peaks at 0.45$\pm$0.28 $R_{\rm bar}$, where $R_{\rm bar}$ is the bar radius determined by maximum ellipticity.
Galaxies with a B/P bulge show enhanced stellar radial motions compared to galaxies without a B/P bulge do. The maximum ratio of non-circular to circular velocity occurs at 1.02$\pm 0.53 R_{\rm BP}$ on average. 
However, this could be because galaxies with a B/P feature in our sample also have strong bars.
More samples with both stellar and gas kinematics, as well as numerical simulations, are necessary to construct a comprehensive understanding of the effect of B/P bulges on non-circular motions.

%\begin{acknowledgments}
\vspace{10mm}
%We thank the referee XXXX.
We thank the anonymous referee for their careful review and comments that helped improve this paper.
This work was supported by the National Research Foundation of Korea (NRF), through grants funded by the Korean government (MSIT) (No. 2022R1A4A3031306).
T.K. acknowledges support from the Basic Science Research Program through the National Research Foundation of Korea (NRF) funded by the Ministry of Education (No. RS-2023-00240212). 
D.A.G. is supported by STFC grants ST/T000244/1 and ST/X001075/1.
Y.H.L is supported by Basic Science Research Program through the National Research Foundation of Korea (NRF) funded by the Ministry of Education (No. RS-2023-00249435).
C.L.C. acknowledges support from the Academia Sinica Institute of Astronomy and Astrophysics.
W.T.K. was supported by a grant from the National Research Foundation of Korea (2022R1A2C1004810).
M.K. is supported by the National Research Foundation of Korea (NRF) grant funded by the Korean government (MSIT) (No. RS-2024-00347548).
M.G.P acknowledges the support from the Basic Science Research Program through the National Research Foundation of Korea (NRF) funded by the Ministry of Education (No. 2019R1I1A3A02062242).
%MUSE :
Based on observations collected at the European Southern Observatory under ESO programmes :
095.D-0091, 096.D-0296,	097.D-0408, 0103.D-0440 (PI: Anderson),
096.B-0309 (PI: Carollo), 
096.D-0263 (PI: Lyman), 
099.D-0022 (PI: Galbany), 
0101.D-0748, 0100.D-0341 (PI: Kuncarayakti), and
0103.A-0637	(PI: Husemann).
Raw and reduced data are available at the ESO Science Archive Facility.

\vspace{5mm}
\facilities{VLT:Yepun, KPNO:Mayall (Mosaic-3), Steward:Bok (90Prime), CTIO:Blanco (DECam)}

\appendix

\begin{figure*}
\includegraphics[width=0.47\textwidth]{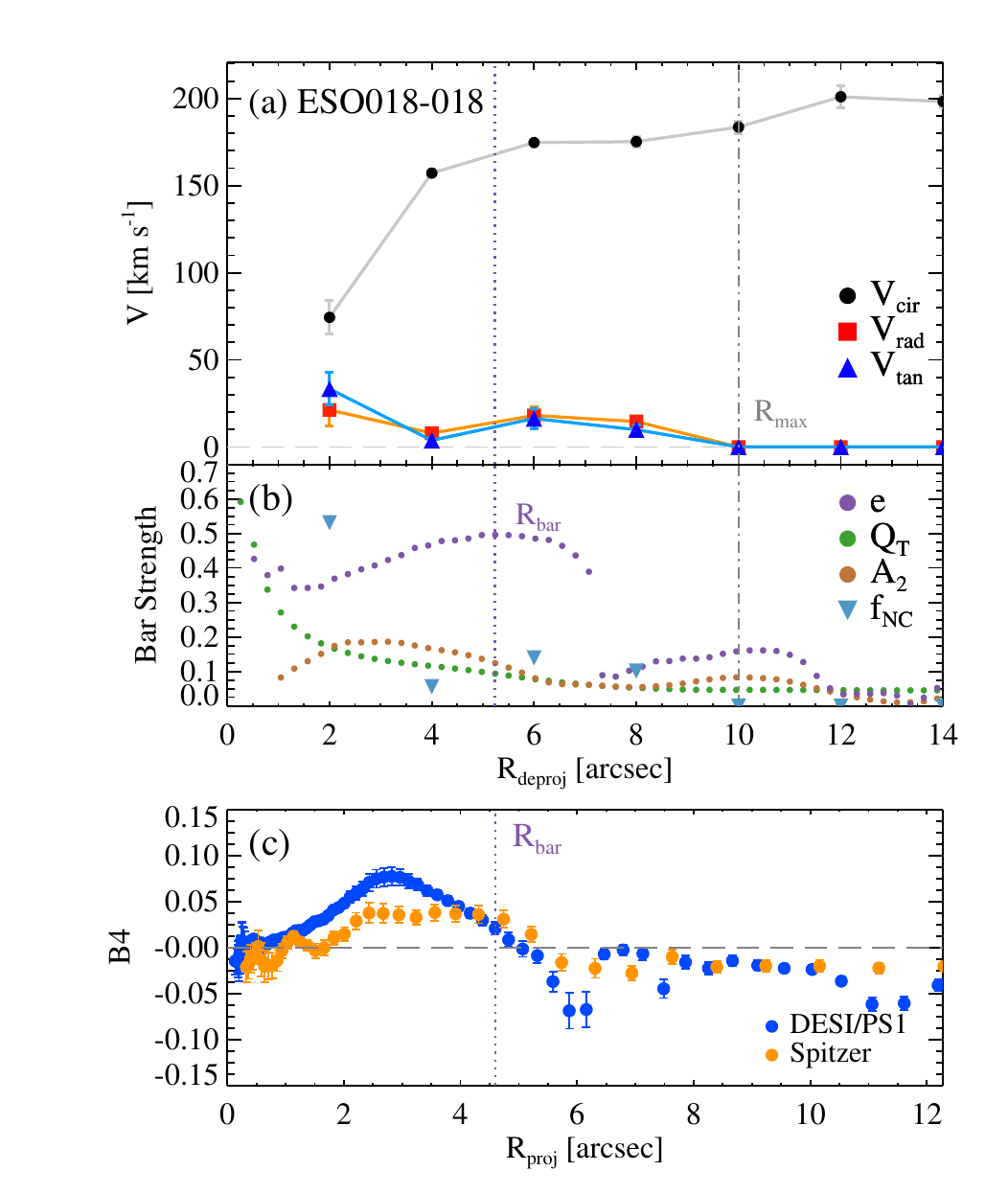}
\includegraphics[width=0.47\textwidth]{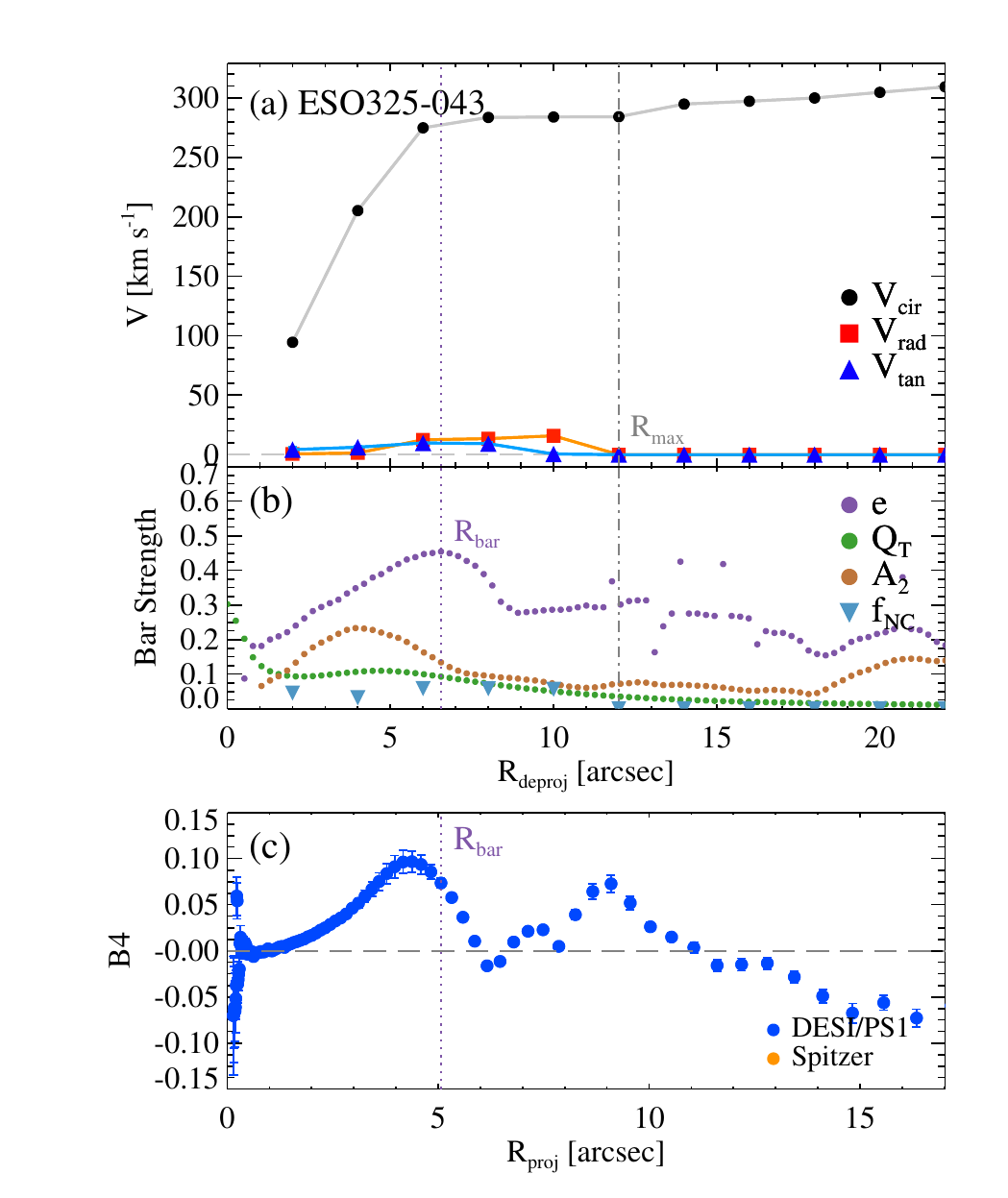}
\includegraphics[width=0.47\textwidth]{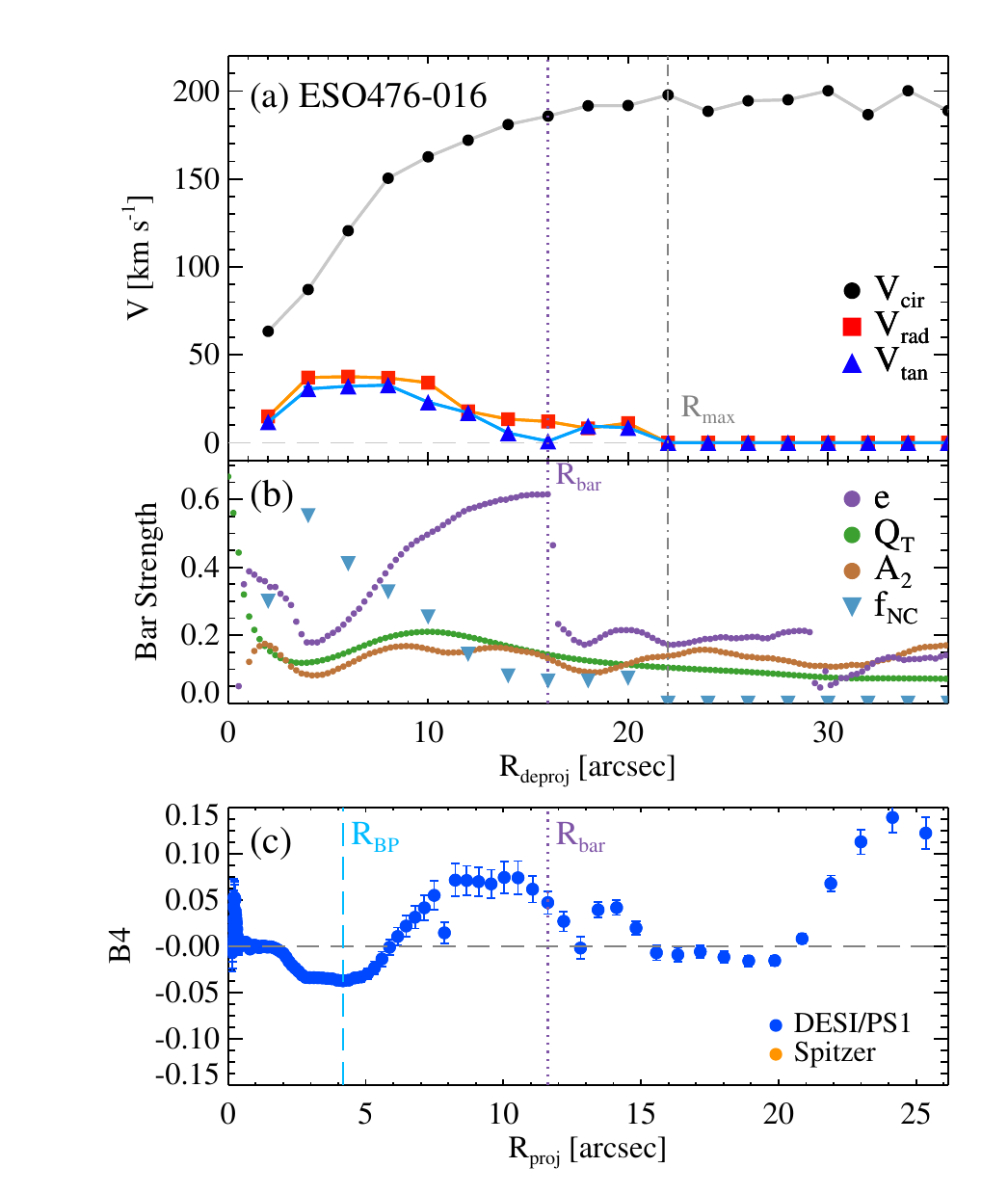}
\includegraphics[width=0.47\textwidth]{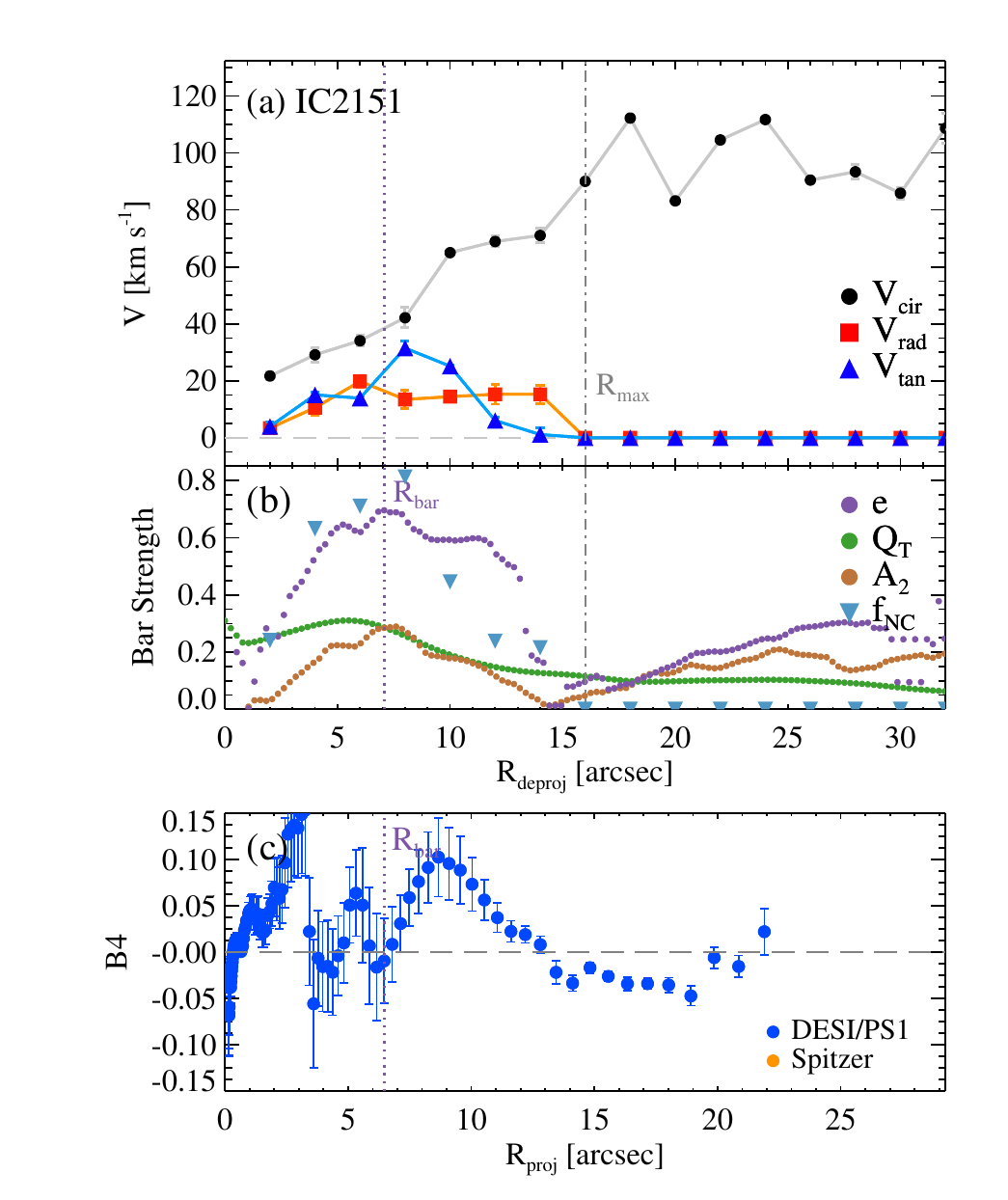}
\caption{Radial profiles of 
(a): $V_{\rm cir},V_{\rm rad}, V_{\rm tan}$ and $f_{\rm NC}$ from fitting the stellar velocity fields with the bisymmetric model \citet{spekkens_07}.
Vertical dashed-dotted line stands for $R_{\max}$ where $V_{\rm rad}$ and $V_{\rm tan}$ (non-circular motions) become zero for more than two consecutive data points.
(b): Ellipticity (e), $Q_b$, and $A_2$. Purple vertical dotted line represents bar radius. All quantities are obtained from deprojected images.
(c): Radial profiles of B4, the boxiness parameter obtained from projected images. The vertical dashed line denotes the size of the B/P $(R_{\rm BP})$ in case where the galaxy exhibits B/P structure.
\label{apfig:rad_profile}
}
\end{figure*}

\begin{figure*}
\includegraphics[width=0.5\textwidth]{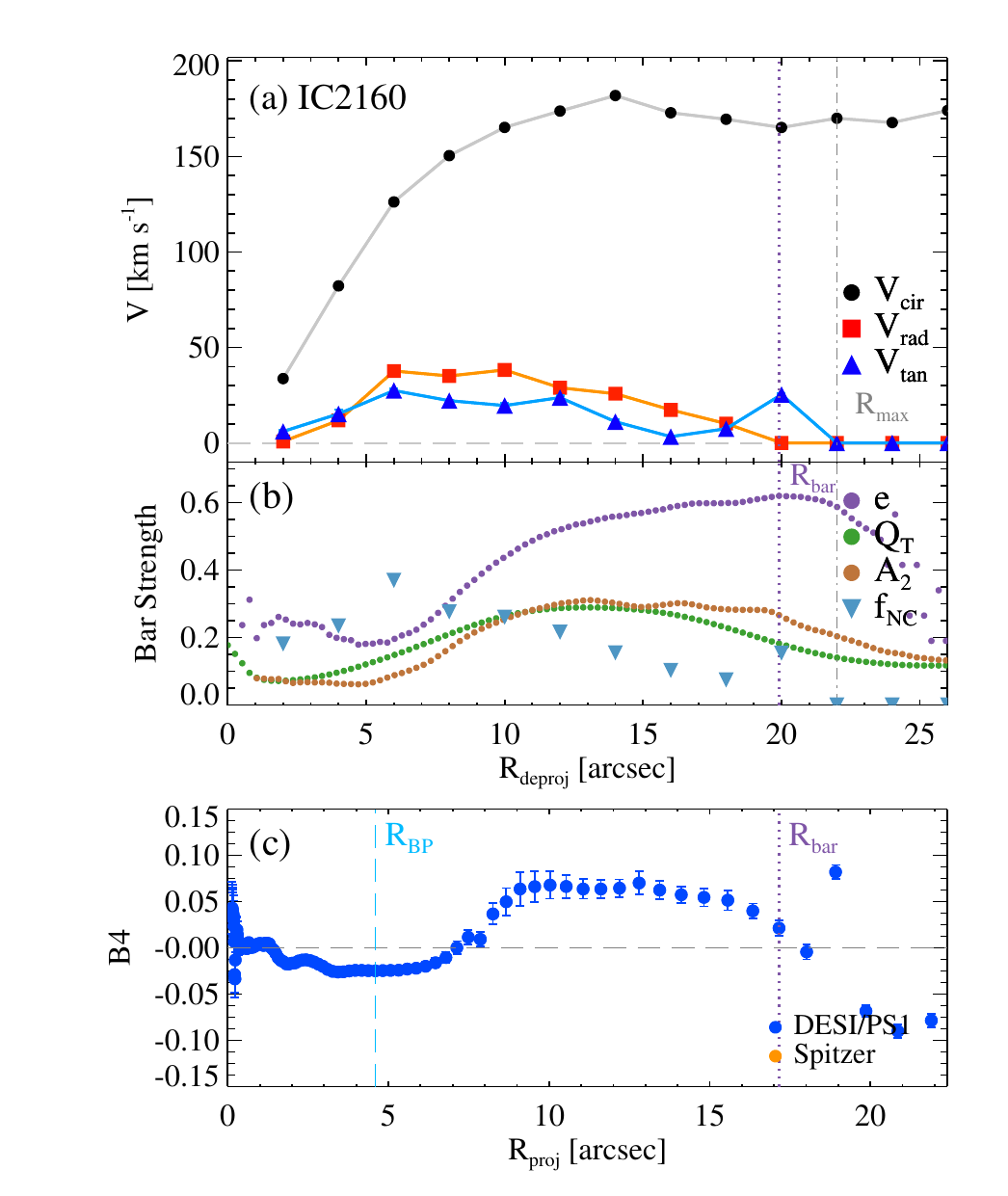}
\includegraphics[width=0.5\textwidth]{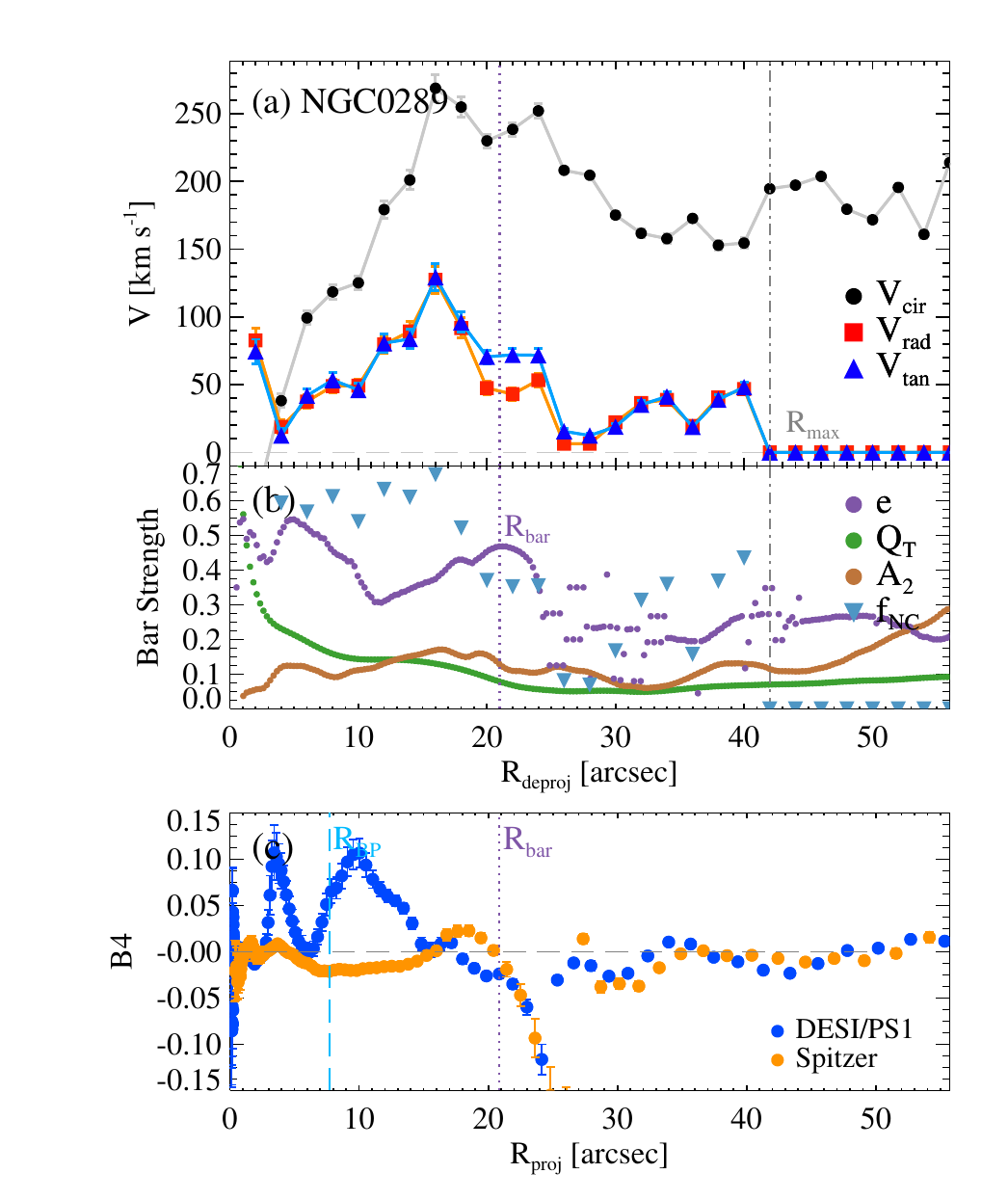}
\includegraphics[width=0.5\textwidth]{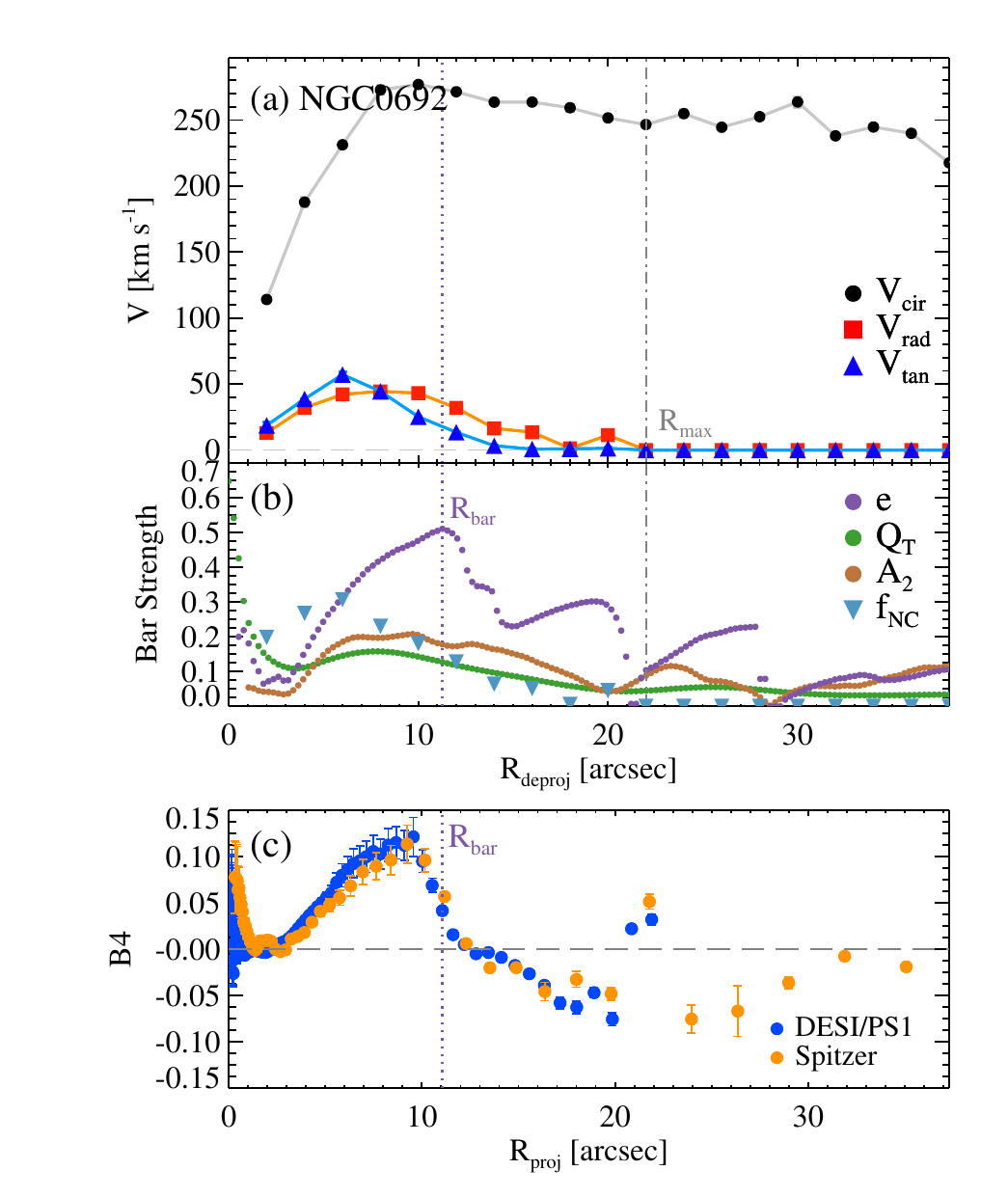}
\includegraphics[width=0.5\textwidth]{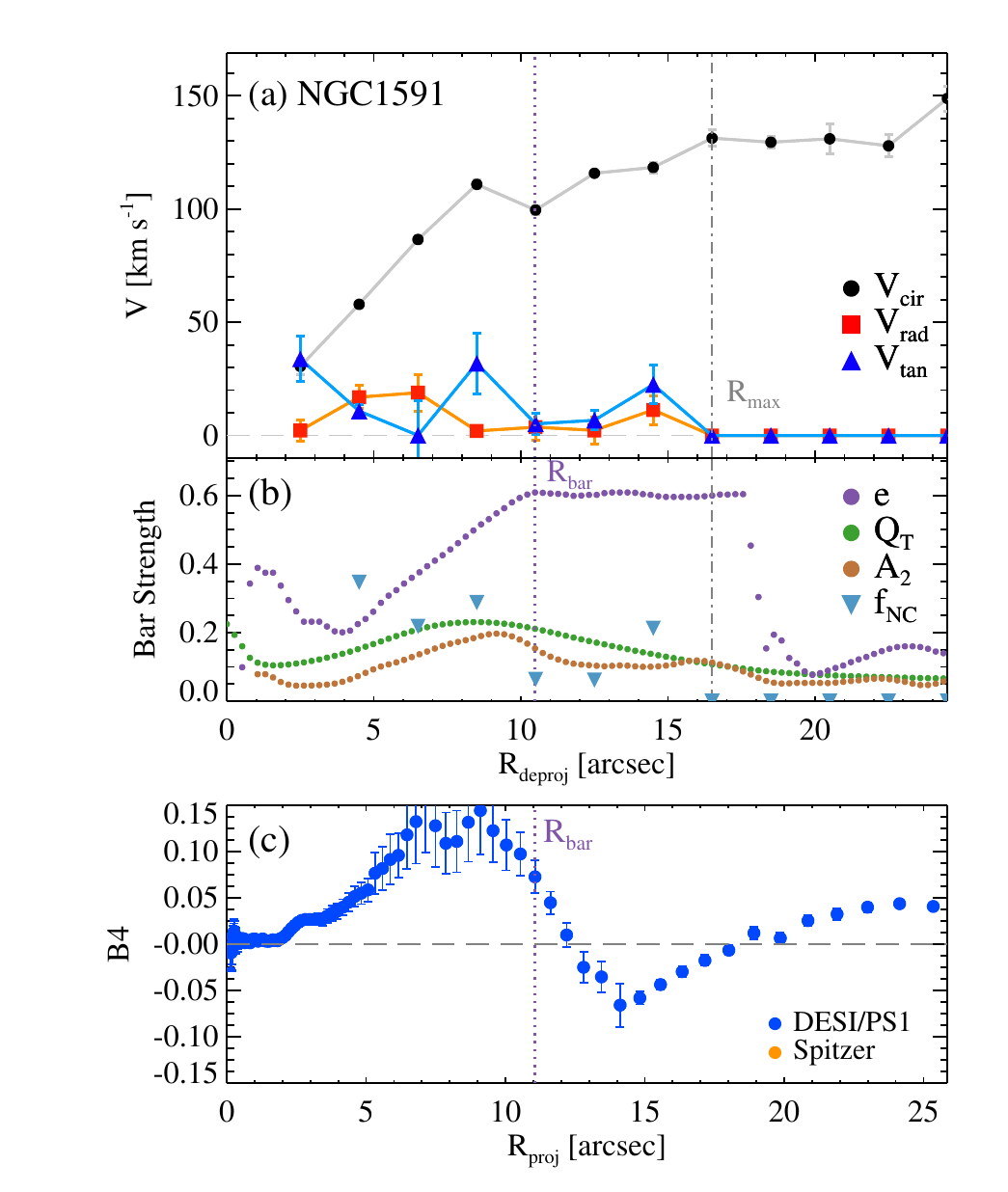}
\caption{Continued}
\end{figure*}

\begin{figure*}
\includegraphics[width=0.5\textwidth]{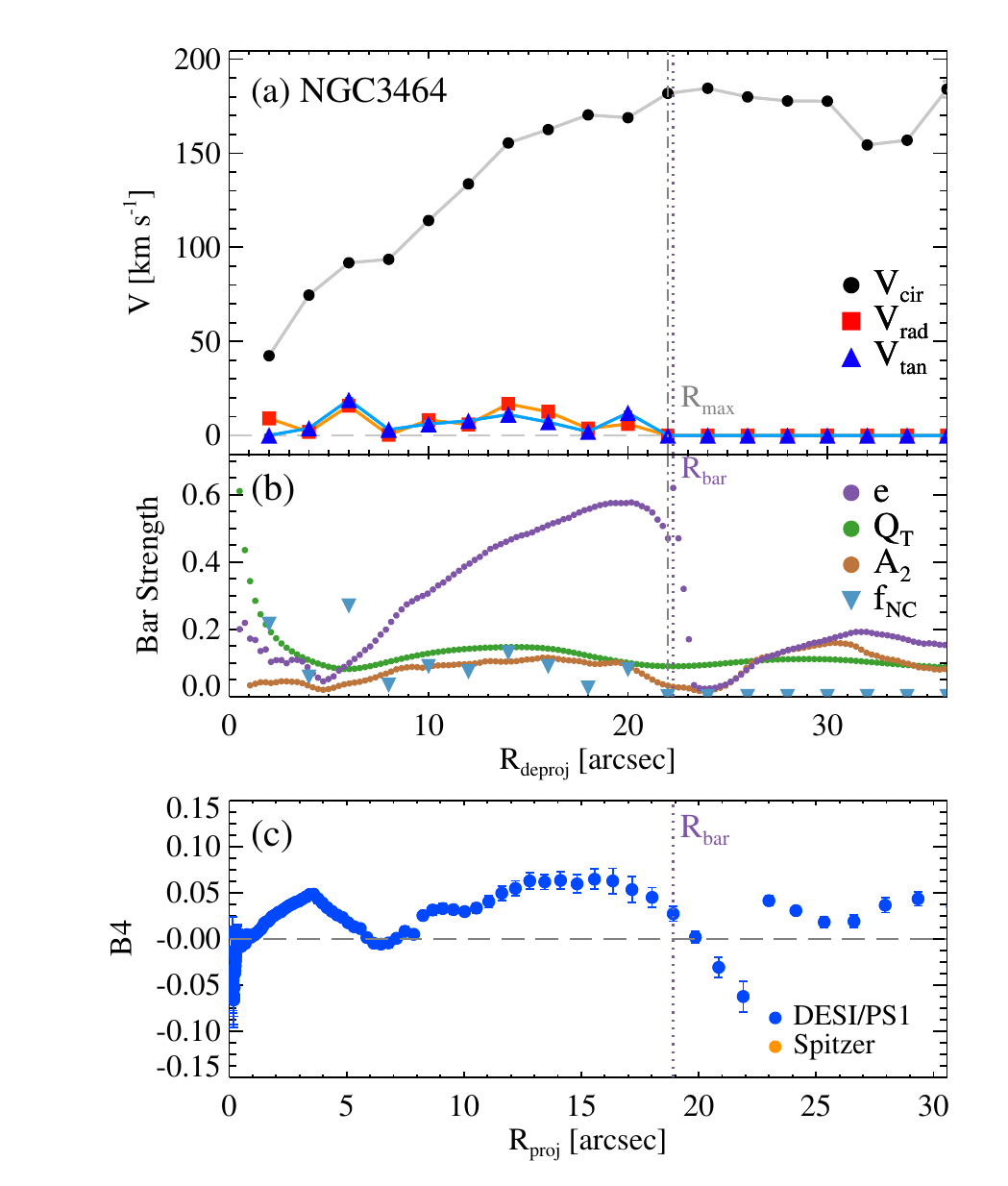}
\includegraphics[width=0.5\textwidth]{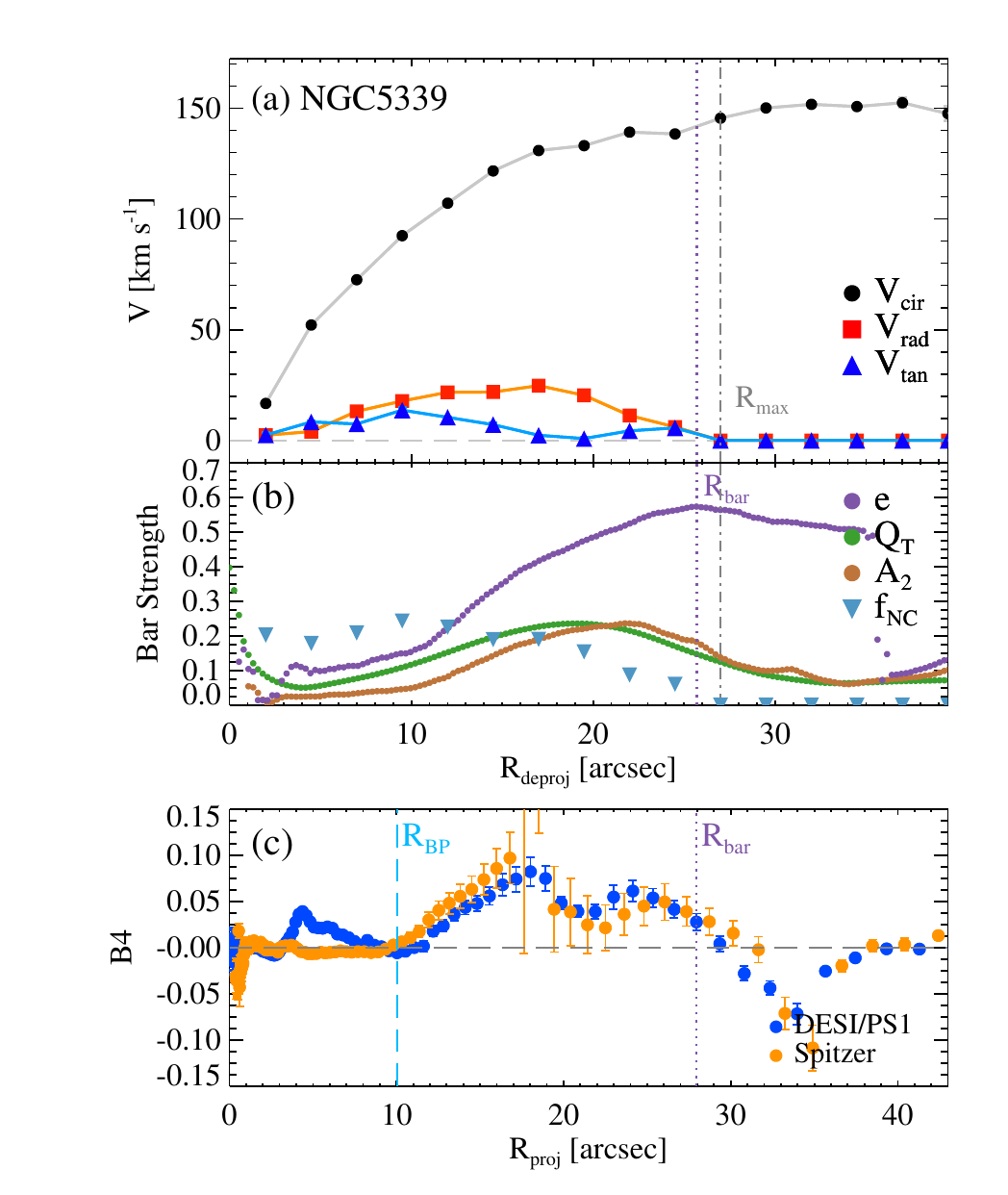}
\includegraphics[width=0.5\textwidth]{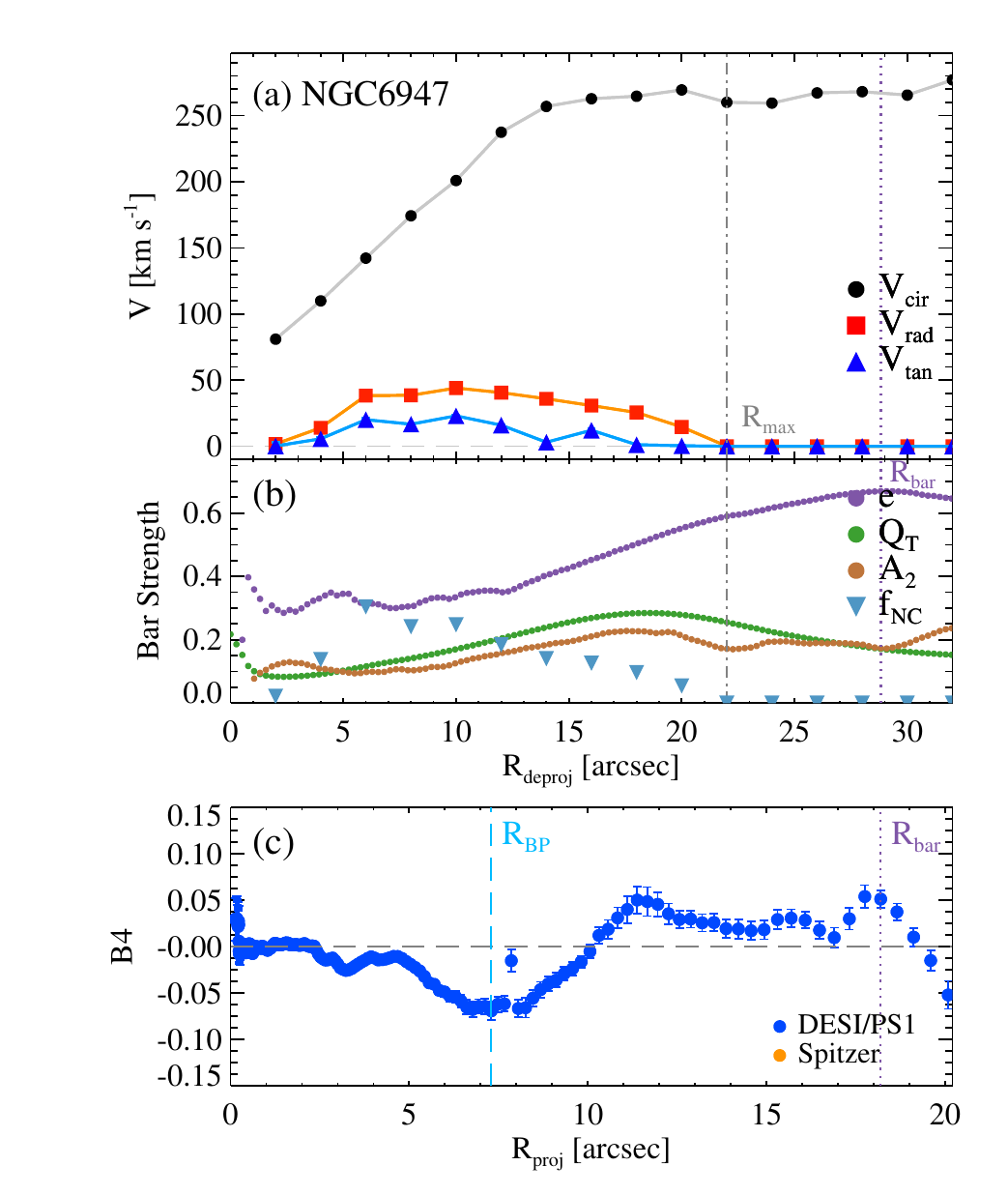}
\includegraphics[width=0.5\textwidth]{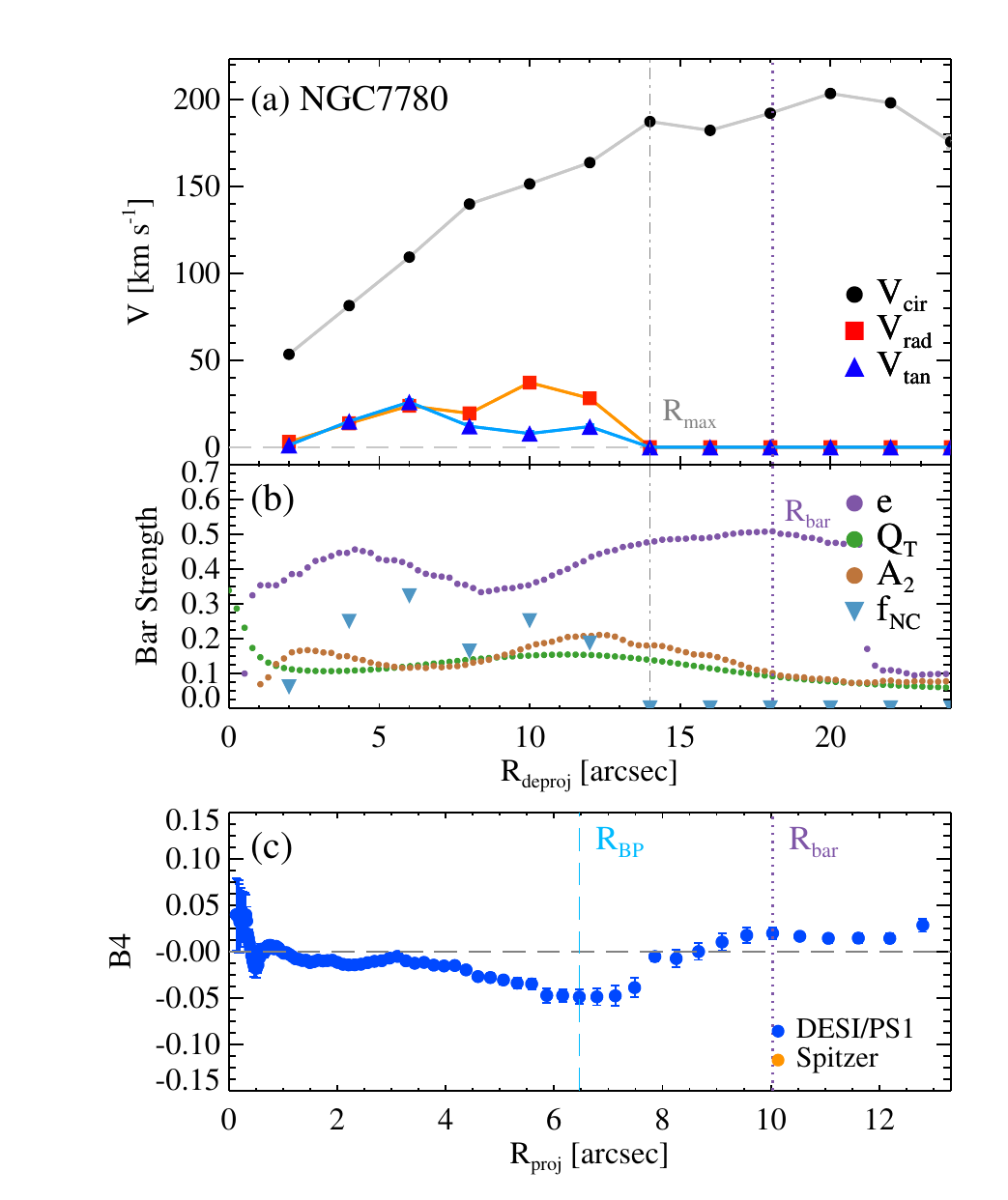}
\caption{Continued}
\end{figure*}

\begin{figure*}[th!]
\includegraphics[width=0.5\textwidth]{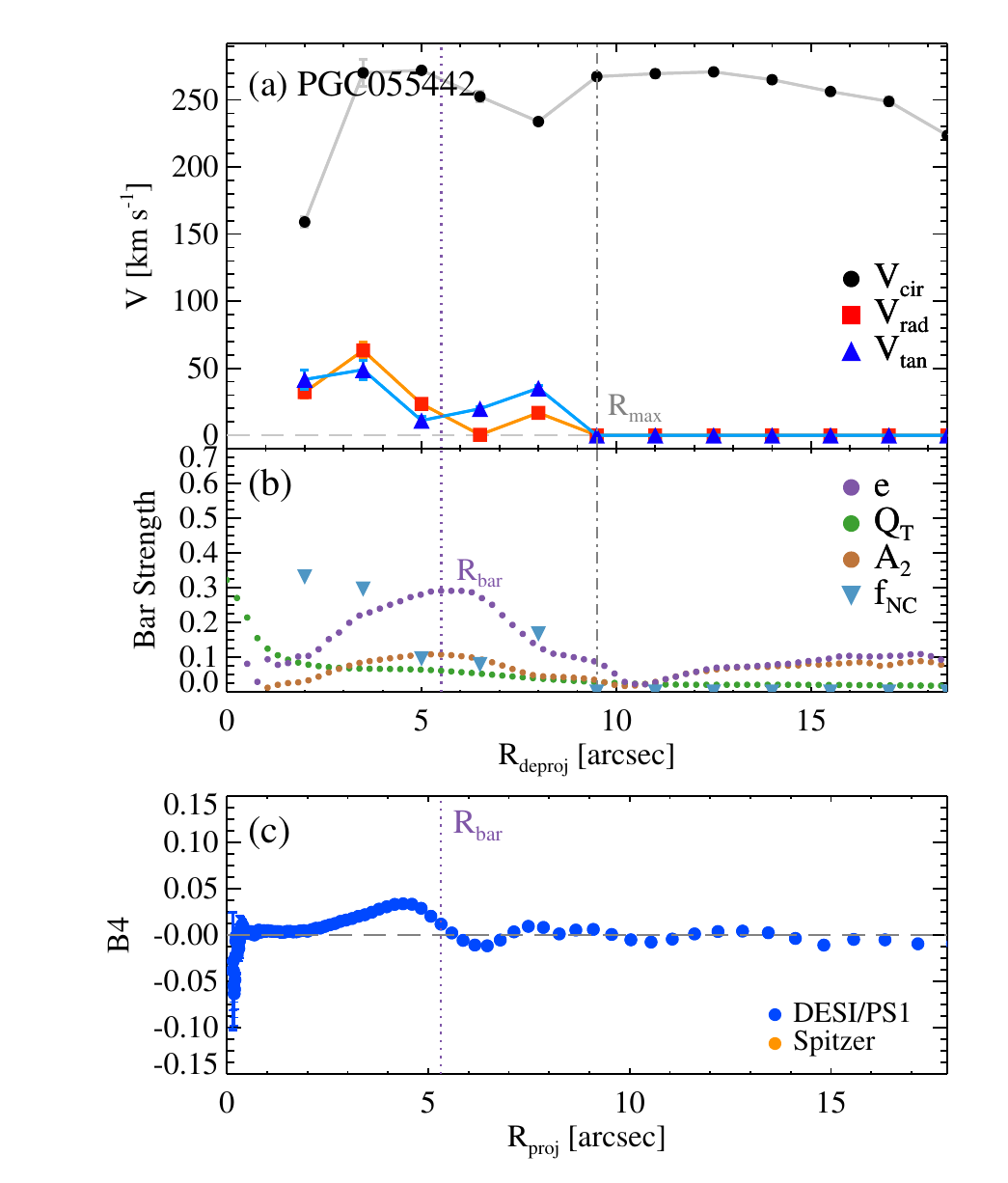}
\includegraphics[width=0.5\textwidth]{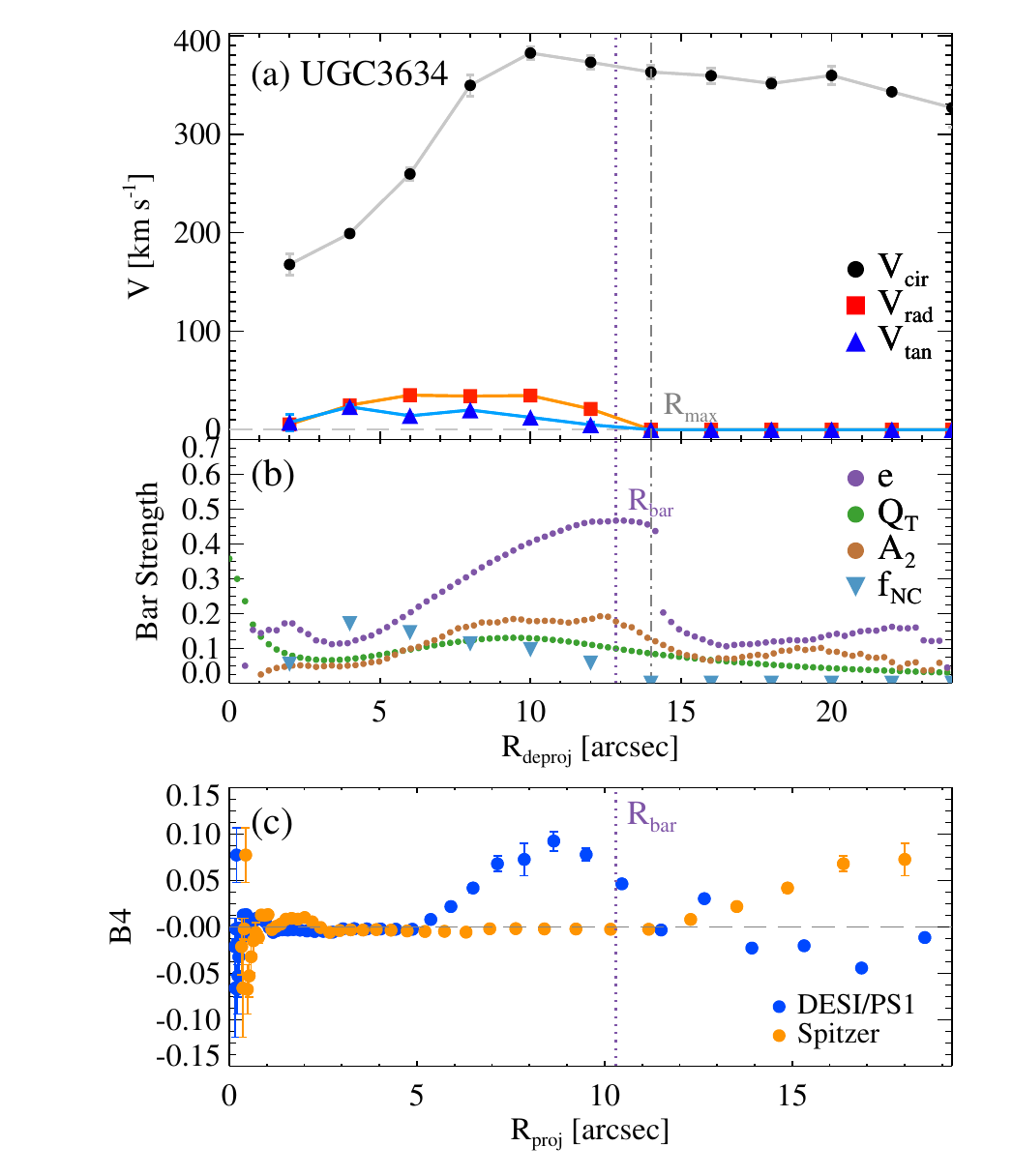}
\caption{Continued}
\end{figure*}

%\section{Boxy/Peanut bulges}\label{sec:app1}
%We present radial profiles of B4 for our sample galaxies in Fig.~\ref{apfig:b4_profile} and  ~\ref{apfig:b4_profile2}. B4 are measured in the r-band (DESI/Pan-STARRS), $\rm K_s$-band (2MASS) and 3.6 $\mu$m (Spitzer) if available.

%\begin{figure*}[ht!]
%\includegraphics[width=\textwidth]{b4_profile1.eps}
%\caption{Profiles of B4 for the whole sample galaxies. If B4 is positive, the isophote at the radius is disky. Conversely, if B4 is negative, the isophote is boxy. Blue points are of the optical data from the DESI or Pan-STARRS, green points are from the 2MASS, and orange points are from the Spitzer, if any. Vertical dahsed lines stand for bar length from the maximum ellipticity, measured  from the projected images.
%\label{apfig:b4_profile}
%}
%\end{figure}

%\begin{figure*}[ht!]
%\includegraphics[width=\textwidth]{b4_profile2.eps}
%\caption{B4 profiles.
%\label{apfig:b4_profile2} 
%}
%\end{figure}



\bibliography{tk24_nc_arxiv}{}
\bibliographystyle{aasjournal}
\end{document}